\documentclass[journal,twocolumn]{IEEEtran}

\usepackage{calligra}
\usepackage{amsfonts,amscd,mathrsfs,amsmath,amsthm,amssymb}
\usepackage{mathtools}
\usepackage{xparse}
\usepackage{graphicx}
\usepackage{float}
\usepackage{pifont}
\usepackage[dvipsnames]{xcolor}
\usepackage{colortbl}
\usepackage{multirow}
\usepackage{enumerate}   
\usepackage{comment}
\usepackage{booktabs}
\usepackage{cancel}
\usepackage{lieart}
\usepackage{mhchem}
\usepackage{listings}
\usepackage{xcolor}
\usepackage{physics}
\usepackage{bm}
\usepackage{bbm}
\usepackage{caption}
\usepackage{lipsum}

\PassOptionsToPackage{hyphens}{url}\usepackage{hyperref}
\hypersetup{
    colorlinks=true,
    linkcolor=magenta!80!black,  
    citecolor=magenta!80!black,
    breaklinks=true,
}
\usepackage[capitalise]{cleveref}

\theoremstyle{theorem}
\newtheorem{definition}{Definition}
\theoremstyle{definition}

\newtheorem*{conjecture*}{Conjecture}
\newtheorem{corollary}{Corollary}
\newtheorem{lemma}{Lemma}
\newtheorem*{lemma*}{Lemma}

\AddToHook{cmd/appendix/before}{%
  \setcounter{equation}{0}%
  \renewcommand{\theequation}{\thesection\arabic{equation}}%
  \setcounter{theorem}{0}%
  \renewcommand{\thetheorem}{\thesection\arabic{theorem}}%
  \setcounter{lemma}{0}%
  \renewcommand{\thelemma}{\thesection\arabic{lemma}}%
}

\newcommand\numberthis{\addtocounter{equation}{1}\tag{\theequation}}

\DeclareMathAlphabet{\mathsfit}{T1}{\sfdefault}{\mddefault}{\sldefault}
\SetMathAlphabet{\mathsfit}{bold}{T1}{\sfdefault}{\bfdefault}{\sldefault}

\newcommand{\ZZ}{\mathbb{Z}}	%Integers
\newcommand{\CC}{\mathbb{C}}
\renewcommand{\SU}{\mathrm{SU}}

\renewcommand{\U}{\mathrm{U}}

\renewcommand{\SO}{\mathrm{SO}}

\newcommand{\smallminus}{\text{-}}

\renewcommand{\G}{\mathcal{G}}

\newcommand{\HH}{\mathcal{H}}

\newcommand{\Ad}{\textsf{Ad}}

\renewcommand{\C}{\mathcal{C}} 
\newcommand{\Cint}{\mathcal{I}}
\newcommand{\emb}{\mathcal{U}}

\renewcommand{\L}{\mathscr{L}} 
\newcommand{\GG}{\mathsf{G}}

\newcommand{\lcode}{\{\mkern-4mu\{}
\newcommand{\rcode}{\}\mkern-4mu\}}

\begin{document}

\title{Intrinsic Quantum Codes}

\author{Eric~Kubischta, Ian~Teixeira

\thanks{Both authors contributed equally to this work.}
\thanks{Eric Kubischta is affiliated with the Department of Mathematics, Florida State University, Tallahassee, FL 32306.}
\thanks{Ian Teixeira is affiliated with the Department of Mathematics, University of California, San Diego, CA 92093}% <-this % stops a space
}

\maketitle

\begin{abstract}
We introduce an intrinsic formulation of quantum error correction based on
representation theory, in which error-protection structure is attached directly
to a unitary group representation, rather than to a particular embedding into a
larger Hilbert space. In this framework, error models are organized by the
isotypic decomposition of the conjugation representation on $\L(V)$. Our main result, the \emph{Schur bootstrap}, shows that if an intrinsic code
satisfies the Knill--Laflamme conditions on a given symmetry sector, then the
same error-protection relations hold for every extrinsic realization obtained
from a $G$-equivariant isometric embedding of the representation into a larger
Hilbert space. Thus a single intrinsic verification certifies error correction
across an entire family of physical realizations.

We further introduce an intrinsic notion of distance, called depth, defined by
adjoint order. For fundamental multi-qudit systems this recovers the
conventional code distance, while for more general representations it refines
the usual weight-based notion. We also prove an intrinsic Eastin--Knill
theorem: any intrinsic code of depth at least two admits only a discrete
logical symmetry group, with the obstruction to continuous covariant gates
arising from the adjoint action rather than from tensor-product structure.

Three worked examples illustrate the framework. The minimal code
$\lcode\irrep{5},2,2\rcode_{\SU(2)}$ unifies the four-qubit
permutation-invariant code and the Chuang--Leung--Yamamoto bosonic code as
realizations of a single intrinsic object. The code $\lcode\irrep{14},2,3\rcode_{\SU(2)}$ yields exponentially large
families of distance-$3$ codes admitting a transversal single-qubit Clifford
gate set. Finally, we construct a $\lcode\irrep{27},5,2\rcode_{\SU(3)}$ code showing that
the framework extends beyond $\SU(2)$, yielding qutrit realizations with
logical symmetry group~$A_6$.
\end{abstract}

\begin{IEEEkeywords}
quantum error correction, representation theory, equivariant codes, Eastin–Knill theorem, transversal gates
\end{IEEEkeywords}

\section{Introduction}

Quantum error correction is traditionally formulated by specifying a code subspace inside a chosen physical Hilbert space and analyzing its protection against a prescribed set of error operators acting on that space. While this approach has produced many powerful constructions, it ties the description of a code to a particular physical realization. In this work we show that the essential error-correction structure can instead be formulated intrinsically, as a property of a group representation itself, independent of the Hilbert space in which that representation is embedded.

The central premise of this work is that quantum error correction admits a
complementary intrinsic formulation.
We model a quantum code as a subspace $\Cint$ of a finite-dimensional unitary
representation $V$ of a symmetry group $G$.
In this setting, error operators are organized by their transformation
properties under the conjugation action of $G$, and error-protection
conditions are expressed directly in representation-theoretic terms.
An \emph{intrinsic quantum code} is therefore not a new physical code, but an
abstract object encoding the error-correction structure common to an entire
family of realizations.

This perspective is closely analogous to the distinction between \emph{intrinsic} and
\emph{extrinsic} geometry.
A Riemannian manifold possesses curvature and geodesic distance determined entirely by
its internal metric, independent of how it is embedded into a higher-dimensional space.
Gauss's \emph{Theorema Egregium}~\cite{gauss1828disquisitiones} established that
Gaussian curvature is an intrinsic invariant, unchanged under isometric embeddings.
In a similar spirit, we show that the defining error protection relations of an
intrinsic quantum code are invariants of the representation itself, independent of the
ambient Hilbert space in which that representation occurs as a subrepresentation.

At a purely linear-algebraic level, any two finite-dimensional code spaces of the same
dimension are isomorphic.
What is nontrivial in quantum error correction is not the existence of linear maps
between code spaces, but the preservation of the algebraic relations that characterize
error protection.
In particular, a quantum code is defined not only by its code space but by how error
operators act relative to that space, as captured by the Knill--Laflamme (KL)
conditions.
Our focus is on identifying when this error correction structure is invariant under
different embeddings of the same representation.
The intrinsic framework isolates precisely this structure and provides criteria under
which it is preserved.

Concretely, the operator algebra $\L(V)$ decomposes under the conjugation action of $G$
as
\begin{equation}\label{eq:intro-isotypic-decomp}
  \L(V)\;\cong\; \bigoplus_{\xi} W_{\xi},
\end{equation}
where $W_{\xi}$ is the $\xi$-isotypic subspace for an irrep $\xi$ of $G$.
Given a subspace $\Cint\subset V$ with orthogonal projector $\Omega$, we say $\Cint$
satisfies the \emph{intrinsic KL condition} for an isotypic sector $W_\xi$ if
\begin{equation}\label{eq:intro-intrinsic-KL}
  \Omega\,F\,\Omega \;=\; c_F\,\Omega
  \qquad\text{for all }F\in W_{\xi}.
\end{equation}
The condition is imposed on the full isotypic subspace $W_\xi$ rather than on
individual irreducible summands within it.
This is essential: when $W_\xi$ has multiplicity $m_\xi > 1$, there is no canonical
way to single out a preferred copy of $\xi$ inside $W_\xi$, and the Schur bootstrap
requires the KL condition to hold across all of $W_\xi$ in order to control which
part of $W_\xi$ the compression map can reach.

Our main technical result, which we call the \emph{Schur bootstrap}, makes the
embedding-invariance statement precise.
Given an intrinsic quantum code satisfying symmetry-resolved KL conditions, we prove
that every $G$-equivariant isometric embedding of $V$ into a larger Hilbert space
$\HH$ yields an extrinsic quantum code with identical error protection properties.
The proof is a direct consequence of Schur's lemma: the compression map
$\Phi(E)=\emb^\dagger E\,\emb$ is $G$-equivariant, so each extrinsic isotypic
sector maps into the corresponding intrinsic sector, and intrinsic KL conditions
propagate automatically to the embedded code.

We further introduce a representation-theoretic notion of code distance.
For $G=\SU(q)$, irreps are organized by their first appearance in tensor powers of
the adjoint representation $\Ad$, inducing an \emph{adjoint order} on irreps and a
corresponding notion of \emph{depth} (intrinsic distance) for an intrinsic code.
For standard multi-qudit codes built from the fundamental representation, adjoint
order coincides with error weight, and depth reduces exactly to conventional code
distance.
For higher or heterogeneous local representations, weight and adjoint order diverge,
and depth provides the appropriate organizing principle.
By the Schur bootstrap, intrinsic depth is preserved under all equivariant embeddings.

The intrinsic viewpoint also clarifies fundamental constraints on fault-tolerant
logical gates.
Defining the logical symmetry group as the quotient $\GG:=\G_{\mathrm{inv}}/\G_0$,
we prove an \emph{intrinsic Eastin--Knill theorem}: if an intrinsic code has depth at
least $2$, then $\GG$ is discrete --- and finite when $G$ is compact.
Combined with the Schur bootstrap, this recovers standard Eastin--Knill-type
obstructions for all extrinsic realizations while pinpointing their
representation-theoretic origin.

We illustrate the framework through several worked examples.
The minimal intrinsic code $\lcode\irrep{5},2,2\rcode_{\SU(2)}$ underlies the
four-qubit permutation-invariant code, the Chuang--Leung--Yamamoto bosonic code,
continuous families of multi-qubit codes, two-qutrit realizations, and molecular
codes.
A second example, $\lcode\irrep{14},2,3\rcode_{\SU(2)}$, yields distance-$3$
permutation-invariant codes with a transversal single-qubit Clifford gate set, as well
as exponentially large families of multi-qubit and heterogeneous multi-qudit
realizations all inheriting the same protection and gate structure. A third example for $\SU(3)$ constructs a distance-$2$ intrinsic code in the irrep $\irrep{27}$ with logical symmetry group $A_6$, yielding both $6$-qutrit and $4$-qutrit realizations.

Beyond conceptual unification, the intrinsic framework provides a practical advantage
for code design: KL conditions need only be verified in the typically much smaller
representation space $V$, and the Schur bootstrap then transfers these guarantees to
all equivariant realizations in arbitrarily large ambient Hilbert spaces.
For multi-qudit codes, where intrinsic and conventional distance coincide, this yields
a computationally efficient environment for code design and analysis.
More broadly, intrinsic quantum codes recast the search for fault-tolerant schemes as
a representation-theoretic problem, complementing recent symmetry-based constructions
that exploit $G$-covariance to obtain structured transversal gate
sets~\cite{gross1,CovariantCatPRL,TessellationCodesPRL,us1,us3}.

\section{Intuition: one intrinsic code, many realizations}\label{sec:intuition}

Before developing the general formalism, we sketch a worked example that captures
the basic message of this paper.
Consider two codes that appear, at first sight, to be entirely unrelated.

The first is the four-qubit permutation-invariant
code~\cite{RuskaiPRL,2004permutation,ouyangPI,Hagiwara2020FourQubitDeletion,Nakayama2020SingleDeletion,Ouyang2021PermutationInvariant,422codePRL},
a $((4,2,2))$ code encoding a single logical qubit with codewords
\begin{align}
  \ket{\bar{0}} &= \tfrac{1}{\sqrt{2}}(\ket{0000}+\ket{1111}), \\
  \ket{\bar{1}} &= \tfrac{1}{\sqrt{6}}\sum_{|x|=2}\ket{x}.
\end{align}
The second is the Chuang--Leung--Yamamoto (CLY) bosonic
code~\cite{CLYcode}, encoding a single logical qubit into two bosonic modes
at fixed total excitation $n_1+n_2=4$, with codewords
\begin{align}
  \ket{\bar{0}} &= \tfrac{1}{\sqrt{2}}(\ket{0,4}+\ket{4,0}), \\
  \ket{\bar{1}} &= \ket{2,2}.
\end{align}
One is a discrete-variable code on a tensor-product Hilbert space; the other is a
bosonic code on a constant-excitation subspace of Fock space.
Even if one notices a superficial correspondence between Hamming-weight subspaces and
constant-excitation subspaces, there is no \emph{a priori} reason to expect their
error-correction properties, covariant gate sets, or Eastin--Knill constraints to be
connected.

The central claim of this paper is that both codes are manifestations of the same
underlying object.
Let $G=\SU(2)$ and let $V=\irrep{5}$ be the spin-$2$ irrep with weight basis
$\{\ket{w}\}_{w=0}^{4}$.
Consider the two-dimensional subspace
\begin{equation}\label{eq:intuition-intrinsic-code}
  \Cint := \mathrm{span}\!\left\{
    \tfrac{1}{\sqrt{2}}(\ket{0}+\ket{4}),\;\ket{2}
  \right\} \subset V,
\end{equation}
with projector $\Omega$.
A direct computation shows $\Omega J_i \Omega = 0$ for $i\in\{x,y,z\}$.
Equivalently, $\Cint$ satisfies the Knill--Laflamme condition for all intrinsic errors
transforming in the adjoint representation $\Ad\cong\irrep{3}$ (and the trivial
sector), so $\Cint$ has intrinsic distance (depth) $\mathsf{d}=2$.

This single intrinsic code has extrinsic realizations in both platforms above.
For four qubits with the diagonal $\SU(2)$ action, the permutation-invariant subspace
$\mathrm{Sym}^4(\CC^2)\subset(\CC^2)^{\otimes 4}$ is precisely the unique copy of the spin $2$ irrep
$\irrep{5}$, with $\ket{w}$ identified with Dicke states of Hamming weight $w$.
Embedding $V$ into $\mathrm{Sym}^4(\CC^2)$ produces the $((4,2,2))$ code above.
Likewise, for two bosonic modes the fixed-excitation subspace $n_1+n_2=4$ carries
the spin-$2$ irrep under the Schwinger map, and the embedding
$\ket{w}\mapsto\ket{w,4-w}$ yields the CLY code.

The unifying point is not a coincidence of basis vectors but a symmetry principle:
both embeddings are $\SU(2)$-equivariant isometries.
Our \emph{Schur bootstrap} shows that symmetry-resolved KL relations
verified intrinsically---here, for the $\irrep{1}$ and $\irrep{3}$ sectors---propagate
automatically to \emph{every} extrinsic realization obtained from a $G$-equivariant
embedding, with no additional computation required.

\section{Main Results}

\subsection{Preliminaries}

Let $V$ be a finite-dimensional unitary representation of a group $G$.
We call $V$ the \emph{intrinsic Hilbert space} and refer to operators in $\L(V)$ as
\emph{intrinsic errors}.
Under the conjugation action of $G$, the operator algebra decomposes into isotypic
subspaces as
\begin{equation}
  \L(V) \;\simeq\; V \otimes V^* \;=\; \bigoplus_\xi W_\xi,
\end{equation}
where $W_\xi \simeq \xi \otimes \CC^{m_\xi}$ is the $\xi$-isotypic subspace
corresponding to the irrep $\xi$ of $G$, occurring with multiplicity $m_\xi$.
We refer to~\cite{fultonharris,Georgi} for background on representation theory.

For example, suppose $V = \irrep{15}$ (Dynkin label $\dynkin{2,1}$) for $G = \SU(3)$.
Then
\begin{equation}
  \L(V) \;=\; \irrep{1} \oplus (2)\irrep{8} \oplus \irrep{10} \oplus
  \irrepbar{10} \oplus (2)\irrep{27} \oplus \irrep{35} \oplus \irrepbar{35}
  \oplus \irrep{64}.
\end{equation}
The $\irrep{8}$-isotypic subspace is $W_{\irrep{8}} = \irrep{8} \otimes \CC^2$,
a $16$-dimensional space carrying two copies of $\irrep{8}$.
Although one can choose a basis splitting $W_{\irrep{8}}$ into two orthogonal copies
of $\irrep{8}$, no such splitting is canonical.

\subsection{Intrinsic Codes}

Let $\Cint \subset V$ be a subspace with orthogonal projector $\Omega: V \to V$, and
let $\mathcal{E}$ be a set of irreps of $G$.
We call $\Cint$ an \emph{intrinsic quantum code} for the error set $\mathcal{E}$ if
the Knill--Laflamme (KL) condition
\begin{equation}
  \Omega\, F\, \Omega \;=\; c_F\, \Omega
\end{equation}
holds for all $F \in W_\xi$ and all $\xi \in \mathcal{E}$.

\textit{Remark.} The KL condition is required to hold on the full isotypic subspace $W_\xi$, not merely
on a single copy of $\xi$ within it.
This is essential: when $m_\xi > 1$, there is no canonical decomposition of $W_\xi$
into irreducible summands, so any condition referring to a specific copy of $\xi$
would be basis-dependent.
Requiring protection across the entire isotypic sector is therefore the natural,
basis-independent formulation.

Note that an intrinsic quantum code is an abstract mathematical object, not a physical code.
It captures the error-correction structure common to an entire family of physical
realizations, each obtained by embedding $V$ into a suitable ambient Hilbert space.

\subsection{Extrinsic Codes}

Let $\HH$ be a unitary $G$-representation, which we call the \emph{extrinsic} or
\emph{physical Hilbert space}.
Operators in $\L(\HH)$ are \emph{extrinsic errors}, and decompose as
\begin{equation}
  \L(\HH) \;\simeq\; \HH \otimes \HH^* \;=\; \bigoplus_\xi K_\xi,
\end{equation}
where $K_\xi$ is the $\xi$-isotypic subspace of $\L(\HH)$.

Let $\emb: V \hookrightarrow \HH$ be a $G$-equivariant isometric embedding
(i.e., $\emb^\dagger \emb = \mathbf{1}_V$ and $\emb \circ \rho_V(g) = \rho_\HH(g)
\circ \emb$ for all $g \in G$).
The image
\begin{equation}
  \C \;:=\; \emb(\Cint) \;\subset\; \HH
\end{equation}
is an \emph{extrinsic quantum code}.
Its orthogonal projector $\Pi: \HH \to \HH$ satisfies
\begin{equation}
  \Pi \;=\; \emb\, \Omega\, \emb^\dagger,
\end{equation}
where $\Omega$ is the projector onto the intrinsic code $\Cint$.
The embedding $\emb$ serves as a dictionary translating the intrinsic code into a
concrete physical realization in $\HH$.

\subsection{Adjoint Order and Intrinsic Distance}

In conventional quantum error correction, the distance of a code measures the minimum
weight of a logical error operator: a distance-$d$ code protects against all errors
acting nontrivially on fewer than $d$ subsystems.
In the intrinsic framework, errors are classified instead by how they arise from tensor
powers of the adjoint representation $\Ad$ of $G = \SU(q)$, yielding a
representation-theoretic notion of distance that recovers the conventional one for
fundamental qudits and refines it beyond that setting.

\begin{definition}[Adjoint order]
For an irrep $\xi$ of $G$, its \textbf{adjoint order} is
\begin{equation}
  \operatorname{ord}(\xi) \;=\; \min\bigl\{\, t \geq 0 : \xi \subset \Ad^{\otimes t} \bigr\},
\end{equation}
with the convention $\Ad^{\otimes 0} := \mathbf{1}$.
For an error operator transforming as $\xi_1 \otimes \cdots \otimes \xi_m$ across
multiple subsystems, its \textbf{total adjoint order} is
$\operatorname{ord}(\xi_1) + \cdots + \operatorname{ord}(\xi_m)$.
\end{definition}

\begin{definition}[Intrinsic distance / depth]
An intrinsic quantum code $\Cint \subset V$ has \textbf{depth} $\mathsf{d}$ if it
satisfies the Knill--Laflamme condition for all error operators of total adjoint order
strictly less than $\mathsf{d}$, and fails for some error of total order $\mathsf{d}$.
\end{definition}

For local subsystems transforming as the fundamental representation $\irrep{q}$ of
$\SU(q)$, the space of local errors decomposes as
\begin{equation}
  \irrep{q} \otimes \irrepbar{q} \;=\; \mathbf{1} \oplus \Ad.
\end{equation}
Every nontrivial local error therefore transforms in $\Ad$ and has adjoint order $1$.
An error acting on $t$ subsystems consequently has total order exactly $t$, so depth
coincides with conventional distance.
For standard multi-qudit codes, the intrinsic framework thus reproduces the usual
weight-based notion of protection, while providing a setting in which KL conditions
can be verified in a smaller representation space and bootstrapped to larger ambient
Hilbert spaces via the Schur bootstrap.

When local subsystems transform in representations other than the fundamental, adjoint
order and error weight can differ, and depth provides the appropriate organizing
principle.

Consider a qutrit transforming as the spin-$1$ irrep $\irrep{3}$ of $\SU(2)$, for
which $\Ad = \irrep{3}$.
The single-subsystem error space decomposes as
\begin{equation}
  \irrep{3} \otimes \irrepbar{3} \;=\; \mathbf{1} \oplus \irrep{3} \oplus \irrep{5}.
\end{equation}
Since $\irrep{5}$ first appears in $\Ad^{\otimes 2}$, it has adjoint order $2$.
A generic single-subsystem error therefore contains components of orders $0$, $1$,
and $2$ simultaneously, and the total adjoint orders for two-subsystem errors are:
\begin{equation}
\begin{aligned}
  \operatorname{ord}(\mathbf{1} \otimes \mathbf{1}) &= 0, \\
  \operatorname{ord}(\irrep{3} \otimes \mathbf{1}) \;=\;
  \operatorname{ord}(\mathbf{1} \otimes \irrep{3}) &= 1, \\
  \operatorname{ord}(\irrep{5} \otimes \mathbf{1}) \;=\;
  \operatorname{ord}(\mathbf{1} \otimes \irrep{5}) \;=\;
  \operatorname{ord}(\irrep{3} \otimes \irrep{3}) &= 2, \\
  \operatorname{ord}(\irrep{3} \otimes \irrep{5}) \;=\;
  \operatorname{ord}(\irrep{5} \otimes \irrep{3}) &= 3, \\
  \operatorname{ord}(\irrep{5} \otimes \irrep{5}) &= 4.
\end{aligned}
\end{equation}
A code of depth $\mathsf{d} = 2$ suppresses all order-$1$ errors---the leading
nontrivial contributions---while permitting uncorrected order-$2$ components at weight
$1$.
Weight-based distance would classify all these as single-subsystem (weight-$1$) errors
and make no such distinction.

The depth notion is thus strictly finer than weight-based distance for higher
representations: it resolves the perturbative structure of errors within a fixed weight
class, identifying which components are suppressed and at what order the first logical
error appears.
By the Schur bootstrap, intrinsic depth is preserved under all $G$-equivariant
isometric embeddings, so this refined protection guarantee transfers automatically to
every extrinsic realization.

\subsection{The Schur Bootstrap}

We now show that the error protection properties of an intrinsic quantum code transfer
to every extrinsic realization.
Let $V$ be a unitary representation of $G$, let $\emb: V \hookrightarrow \HH$ be a
$G$-equivariant isometric embedding, and let $\Omega$ and
$\Pi := \emb\,\Omega\,\emb^\dagger$ be the projectors onto the intrinsic code
$\Cint \subset V$ and the extrinsic code $\C := \emb(\Cint) \subset \HH$,
respectively.

The key tool is the \emph{compression map}
\begin{equation}
  \Phi : \L(\HH) \to \L(V), \qquad \Phi(E) := \emb^\dagger E\, \emb,
\end{equation}
which pulls back extrinsic errors to intrinsic ones.
Because $\emb$ intertwines the $G$-actions on $V$ and $\HH$, the compression map is
$G$-equivariant:
\begin{equation}
  \Phi(g E g^{-1}) = g\,\Phi(E)\,g^{-1} \qquad \text{for all } g \in G.
\end{equation}
By Schur's lemma, a $G$-equivariant linear map must send each isotypic component into
the corresponding isotypic component.
Hence, if $K_\xi \subset \L(\HH)$ and $W_\xi \subset \L(V)$ denote the
$\xi$-isotypic subspaces of extrinsic and intrinsic errors respectively, then
\begin{equation}
  \Phi(K_\xi) \subseteq W_\xi.
\end{equation}

Suppose the intrinsic code $\Cint$ satisfies the Knill--Laflamme condition for all
errors in $W_\xi$.
Then for any $E \in K_\xi$ we have $\Phi(E) \in W_\xi$, and therefore
\begin{equation}
  \Pi E \Pi
  = \emb\,\Omega\,\Phi(E)\,\Omega\,\emb^\dagger
  = c_E\,\emb\,\Omega\,\emb^\dagger
  = c_E\,\Pi,
\end{equation}
where the second equality uses the intrinsic KL condition applied to $\Phi(E) \in W_\xi$.
Thus $\C$ satisfies the KL condition for all extrinsic errors in $K_\xi$.

The argument requires the KL condition to hold for the \emph{entire} isotypic subspace
$W_\xi$.
Since $\Phi$ is only required to satisfy $\Phi(K_\xi) \subseteq W_\xi$, one has no
control over which elements of $W_\xi$ are in the image of $\Phi$; the intrinsic KL
condition must therefore cover all of $W_\xi$ to guarantee that $\Phi(E)$ is detectable
for every $E \in K_\xi$.
This is precisely why intrinsic codes are defined with respect to full isotypic sectors
rather than individual irreducible summands.

\begin{lemma}[Schur Bootstrap]\label{lem:SchurBootstrap}
Let $\Cint \subset V$ be an intrinsic quantum code satisfying the Knill--Laflamme
condition for all errors in an isotypic subspace $W_\xi \subset \L(V)$.
Then for any $G$-equivariant isometric embedding $\emb: V \hookrightarrow \HH$, the
extrinsic code $\C = \emb(\Cint) \subset \HH$ satisfies the Knill--Laflamme condition
for all errors in the corresponding isotypic subspace $K_\xi \subset \L(\HH)$.
\end{lemma}

\begin{corollary}\label{cor:depth-preserved}
Intrinsic depth is preserved under $G$-equivariant isometric embeddings: if $\Cint$
has depth $\mathsf{d}$, then every extrinsic realization $\C = \emb(\Cint)$ also has
depth $\mathsf{d}$.
\end{corollary}

For multi-qudit codes, where intrinsic depth coincides with conventional distance, the
Schur bootstrap provides a practically useful reduction: KL conditions need only be
verified within the intrinsic representation space $V$, which is typically far smaller
than any ambient physical Hilbert space $\HH$, and the resulting distance guarantees
transfer automatically to all equivariant realizations.

\subsection{Covariant Gates and an Intrinsic Eastin--Knill Theorem}

Let $G$ be a Lie group acting unitarily on a finite-dimensional representation $V$, and
let $\Cint \subset V$ be an intrinsic quantum code with projector $\Omega$.
Define the \emph{invariant subgroup}
\begin{equation}
  \G_{\mathrm{inv}} := \{\, g \in G : g\,\Cint = \Cint \,\}
\end{equation}
and the \emph{pointwise stabilizer}
\begin{equation}
  \G_0 := \{\, g \in G : g\ket{\psi} = \ket{\psi}
  \text{ for all } \ket{\psi} \in \Cint \,\}.
\end{equation}
The quotient group
\begin{equation}
  \GG := \G_{\mathrm{inv}}/\G_0
\end{equation}
acts faithfully on $\Cint$ and represents the logical operations induced by the
$G$-action — the intrinsic analogue of the transversal gate group for multi-qudit
codes.
Because the embedding $\emb$ is $G$-equivariant, $\C = \emb(\Cint)$ is also
$\GG$-covariant: if $g \in G$ enacts $\hat{g} \in \GG$ on $\Cint$, then $g$ enacts
the same $\hat{g}$ on $\C$.

The presence of error protection places strong restrictions on $\GG$.

\begin{lemma}[Intrinsic Eastin--Knill]\label{lem:IntrinsicEK}
Let $G$ be a semisimple Lie group and let $\Cint \subset V$ be an intrinsic quantum
code satisfying the Knill--Laflamme condition for all errors transforming in the
adjoint representation $\Ad$ of $G$ --- equivalently, let $\Cint$ have intrinsic
depth $\mathsf{d} \geq 2$.
Then the logical symmetry group $\GG$ is discrete.
If $G$ is compact, then $\GG$ is finite.
\end{lemma}

\begin{proof}
Let $\rho $ denote the representation of $ G $ on $ V $, and let
$d\rho $ be the corresponding derived representation.
By definition of the conjugation action of $G$ on $\L(V)$, one has
\begin{equation}\label{eq:drho-ad-equiv}
  \rho(g)\, d\rho(X)\, \rho(g)^{-1}
  = d\rho(\Ad_g X)
  \qquad
  \text{for all } g \in G,\ X \in \mathfrak g.
\end{equation}
Thus the subspace $d\rho(\mathfrak g)\subseteq \L(V)$ is a $G$-subrepresentation
of $\L(V)$ carrying the adjoint action (more precisely, a quotient of the adjoint
representation if $\rho$ is not faithful).

Since $\Cint$ satisfies the Knill--Laflamme condition on the adjoint sector,
for each $X \in \mathfrak g$ there exists a scalar $\lambda(X)\in\C$ such that
\begin{equation}\label{eq:EK-KL-scalar}
  \Omega\, d\rho(X)\, \Omega
  = \lambda(X)\,\Omega.
\end{equation}
The map $X \mapsto \lambda(X)$ is linear, because both $d\rho$ and compression by
$\Omega$ are linear.
Moreover, \eqref{eq:drho-ad-equiv} and the $G$-invariance of the KL condition imply
that $\lambda$ is $\Ad$-invariant:
\[
\lambda(\Ad_g X)=\lambda(X)
\qquad
\text{for all } g\in G,\ X\in\mathfrak g.
\]
Equivalently, $\lambda$ is a $G$-invariant linear functional on the adjoint
representation.
Because $G$ is semisimple, the adjoint representation has no nonzero trivial
subrepresentation; equivalently, the only $\Ad$-invariant linear functional on
$\mathfrak g$ is the zero functional.
Hence $\lambda \equiv 0$, and therefore
\begin{equation}\label{eq:EK-compression-zero}
  \Omega\, d\rho(X)\, \Omega = 0
  \qquad
  \text{for all } X \in \mathfrak g.
\end{equation}

We now show that the connected component of the invariant subgroup acts trivially
on $\Cint$.
Let $\G_{\mathrm{inv}}^0$ denote the identity component of $\G_{\mathrm{inv}}$, and
let $\mathfrak g_{\mathrm{inv}}$ be its Lie algebra.
Fix $X\in \mathfrak g_{\mathrm{inv}}$ and $\ket{\psi}\in \Cint$.
Since $X$ lies in the Lie algebra of $\G_{\mathrm{inv}}^0$, the curve
$t\mapsto \exp(tX)$ lies in $\G_{\mathrm{inv}}^0\subseteq \G_{\mathrm{inv}}$ for all
$t$ sufficiently small.
By definition of $\G_{\mathrm{inv}}$, each $\rho(\exp(tX))$ preserves $\Cint$, so
\[
\rho(\exp(tX))\ket{\psi}\in \Cint
\qquad
\text{for all sufficiently small } t.
\]
Differentiating at $t=0$ yields
\begin{equation}\label{eq:EK-tangent-in-code}
  d\rho(X)\ket{\psi}\in \Cint.
\end{equation}
On the other hand, applying \eqref{eq:EK-compression-zero} to $X$ gives
\[
\Omega\, d\rho(X)\,\Omega =0.
\]
Since $\Omega\ket{\psi}=\ket{\psi}$ for $\ket{\psi}\in\Cint$, this implies
\[
\Omega\, d\rho(X)\ket{\psi}=0.
\]
But by \eqref{eq:EK-tangent-in-code}, the vector $d\rho(X)\ket{\psi}$ already lies in
$\Cint$, and $\Omega$ acts as the identity on $\Cint$.
Therefore
\[
d\rho(X)\ket{\psi}=0
\qquad
\text{for all } X\in \mathfrak g_{\mathrm{inv}},\ \ket{\psi}\in\Cint.
\]
Hence the induced Lie algebra action of $\mathfrak g_{\mathrm{inv}}$ on $\Cint$ is
trivial.

It follows that every one-parameter subgroup of $\G_{\mathrm{inv}}^0$ acts trivially
on $\Cint$, and therefore the whole connected Lie group $\G_{\mathrm{inv}}^0$ acts
trivially on $\Cint$.
Equivalently,
\begin{equation}\label{eq:EK-G0containsGinv0}
  \G_{\mathrm{inv}}^0 \subseteq \G_0.
\end{equation}

Now $\G_0 \subseteq \G_{\mathrm{inv}}$ by definition, so
\[
\G_{\mathrm{inv}}^0 \subseteq \G_0 \subseteq \G_{\mathrm{inv}}.
\]
Passing to quotients gives
\begin{equation}\label{eq:EK-quotient}
  \GG
  = \frac{\G_{\mathrm{inv}}}{\G_0}
  \cong
  \frac{\G_{\mathrm{inv}}/\G_{\mathrm{inv}}^0}
       {\G_0/\G_{\mathrm{inv}}^0}.
\end{equation}
The group $\G_{\mathrm{inv}}/\G_{\mathrm{inv}}^0$ is the component group of the Lie
group $\G_{\mathrm{inv}}$, hence it is discrete.
Therefore its quotient $\GG$ is also discrete.

If $G$ is compact, then $\G_{\mathrm{inv}}$ is a closed subgroup of a compact group,
hence compact.
Consequently, $\GG$ is a discrete quotient of a compact group, and therefore finite.
\end{proof}

This result shows that the obstruction to continuous covariant logical gates is already
present at the intrinsic level: whenever an intrinsic code has depth $\mathsf{d} \geq 2$,
all connected components of the symmetry group act trivially on the codespace, leaving
at most a discrete group of logical symmetries.
Combined with the Schur bootstrap, this constraint propagates immediately to every
extrinsic realization.
More importantly, it clarifies the conceptual origin of the Eastin--Knill theorem: the
incompatibility between continuous covariance and error correction arises from the
representation-theoretic structure of the adjoint action, rather than from any specific
features of tensor-product Hilbert spaces.

\section{Example: An $\SU(2)$ Intrinsic Code and Its Realizations}
\label{sec:example-su2}

We illustrate the general framework through a single intrinsic code and several of its
extrinsic realizations.
The example demonstrates how one representation-theoretic object simultaneously
determines codes on qubits, qutrits, and bosonic modes, while making the distinction
between depth and conventional distance explicit.

\subsection{An Intrinsic Code in $V = \irrep{5}$}
\label{sec:example-intrinsic}

Let $G = \SU(2)$ and let $V = \irrep{5}$ be the spin-$2$ irrep.
Fix the weight basis $\{\ket{w}\}_{w=0}^{4}$ (with $\ket{w} := \ket{m=w-2}$).
Under the adjoint action, the operator algebra decomposes as
\begin{equation}
  \L(V) \;\cong\; \irrep{5} \otimes \irrepbar{5}
  \;=\; \irrep{1} \oplus \irrep{3} \oplus \irrep{5} \oplus \irrep{7} \oplus \irrep{9}.
  \label{eq:L(V)-su2-5}
\end{equation}
Consider the two-dimensional subspace $\Cint \subset V$ spanned by
\begin{equation}
  \ket{\bar{0}} := \tfrac{1}{\sqrt{2}}\bigl(\ket{0}+\ket{4}\bigr),
  \qquad
  \ket{\bar{1}} := \ket{2}.
  \label{eq:intrinsic-codewords-su2-5}
\end{equation}
A direct computation using the spin-$2$ matrices shows
\begin{equation}
  \Omega J_i \Omega = 0, \qquad i \in \{x,y,z\},
  \label{eq:OmegaJiOmega}
\end{equation}
so $\Omega F \Omega = 0$ for every $F \in \mathfrak{su}(2) \cong \Ad$.
Equivalently, $\Cint$ satisfies the KL condition for all errors in the $\irrep{1}$ and
$\irrep{3}$ isotypic sectors, giving intrinsic depth $\mathsf{d} = 2$.
We denote this intrinsic code by $\lcode\irrep{5}, 2, 2\rcode_{\SU(2)}$.
Explicit matrices and the verification of~\eqref{eq:OmegaJiOmega} are provided in the
Supplemental Material~\cite{supp}.

\subsection{Four Qubits: A Permutation-Invariant Code}
\label{sec:example-4qubit}

Let $\HH = (\CC^2)^{\otimes 4}$ with the diagonal $\SU(2)$ action $g \mapsto
g^{\otimes 4}$.
The decomposition
\begin{equation}
  (\irrep{2})^{\otimes 4} \;\cong\; \irrep{5} \oplus (2)\irrep{3} \oplus (2)\irrep{1}
\end{equation}
contains a unique copy of $\irrep{5}$, namely the permutation-invariant subspace
$\mathrm{Sym}^4(\CC^2)$.
The (unique up to phase) $\SU(2)$-equivariant isometry $\emb: \irrep{5}
\hookrightarrow \HH$ identifies $\ket{w}$ with the normalized Dicke state of Hamming
weight $w$:
\begin{align}
  \ket{0} &\mapsto \ket{0000}, \nonumber\\
  \ket{1} &\mapsto \tfrac{1}{2}(\ket{0001}+\ket{0010}+\ket{0100}+\ket{1000}),
  \nonumber\\
  \ket{2} &\mapsto \tfrac{1}{\sqrt{6}}(\ket{0011}+\ket{0101}+\ket{0110}
            \nonumber \\& \qquad \quad +\ket{1001}+\ket{1010}+\ket{1100}), \nonumber\\
  \ket{3} &\mapsto \tfrac{1}{2}(\ket{1110}+\ket{1101}+\ket{1011}+\ket{0111}),
  \nonumber\\
  \ket{4} &\mapsto \ket{1111}.
\end{align}
The extrinsic code $\C = \emb(\Cint)$ has logical states
\begin{align}
  \ket{\bar{0}} &= \tfrac{1}{\sqrt{2}}(\ket{0000}+\ket{1111}), \nonumber\\
  \ket{\bar{1}} &= \tfrac{1}{\sqrt{6}}(\ket{0011}+\ket{0101}+\ket{0110}
                 \nonumber  \\
                & \quad \qquad +\ket{1001}+\ket{1010}+\ket{1100}),
\end{align}
which is the well-known four-qubit permutation-invariant
code~\cite{RuskaiPRL,2004permutation,ouyangPI,Hagiwara2020FourQubitDeletion,
Nakayama2020SingleDeletion,Ouyang2021PermutationInvariant,422codePRL}.
By the Schur bootstrap (Lemma~\ref{lem:SchurBootstrap}), $\C$ satisfies the KL
condition for all extrinsic errors in the $\irrep{1}$ and $\irrep{3}$ isotypic
subspaces of $\L(\HH)$.
In the standard multi-qubit setting every nontrivial single-qubit error transforms in
$\irrep{3}$, so $\C$ detects all weight-$1$ errors and has conventional distance $2$.

We emphasize that the Schur bootstrap gives a strictly stronger guarantee than the
weight-based statement: $\C$ detects \emph{every} error component lying in the
$\irrep{1}$ or $\irrep{3}$ isotypic sectors, regardless of the weight of the operator.
A two-qubit operator, for instance, may contain both detectable components (in $\irrep{1}$
or $\irrep{3}$) and undetectable ones (in higher irreps), and the Schur bootstrap
certifies detection of the former from representation type alone.

\subsection{A Continuous Family of Six-Qubit Realizations}
\label{sec:example-moduli}

When the target space contains multiple copies of $\irrep{5}$, equivariant embeddings
are no longer unique, and the intrinsic code gives rise to a continuous family of
extrinsic realizations.
For $\HH = (\CC^2)^{\otimes 6}$, the irrep $\irrep{5}$ appears with multiplicity $5$.
Any normalized linear combination $\emb_z = \sum_{i=1}^{5} z_i \emb_i$ with
$[z] \in \CC P^4$ is again an equivariant isometric embedding, defining an extrinsic
code $\C_z := \emb_z(\Cint)$ .
By Lemma~\ref{lem:SchurBootstrap}, every member of this $\CC P^4$ family is a
$((6,2,2))$ code with the same symmetry-resolved error detection guarantees. For an explicit parameterization this family of six qubit codes see the supplemental material \cite{supp}.

% More generally, for even $n \geq 6$, the multiplicity $m_n$ of $\irrep{5}$ in
% $(\irrep{2})^{\otimes n}$ grows as
% \begin{equation}
%   m_n \;=\; \frac{(2+\varphi)^{n/2-1} + (2-\varphi^{-1})^{n/2-1}}{\sqrt{5}}
%   \;\sim\; \mathcal{O}\!\left((2+\varphi)^{n/2}\right),
% \end{equation}
% where $\varphi = (1+\sqrt{5})/2$ is the golden ratio.The intrinsic code $\lcode\irrep{5},2,2\rcode$ therefore admits a $\CC P^{m_n-1}$
% moduli space of extrinsic $((n,2,2))$ realizations, growing exponentially in $n$.

More generally, for even $n \ge 4$, the multiplicity $m_n$ of $\irrep{5}$ in
$(\irrep{2})^{\otimes n}$ is
\begin{equation}
  m_n  = \frac{10}{n+6}\binom{n}{\frac n2-2} \sim \mathcal{O}\!\left(\frac{2^n}{\sqrt{n}}\right). 
\end{equation}
Thus the intrinsic code $\lcode\irrep{5},2,2\rcode$ admits a
$\CC P^{m_n-1}$ moduli space of extrinsic $((n,2,2))$ realizations, growing exponentially in $n$.

\subsection{Bosonic Modes: The CLY Code}
\label{sec:example-bosonic}

For two bosonic modes $(a,b)$, the fixed-excitation subspace $n_1+n_2 = N$ carries
the spin-$j = N/2$ irrep of $\SU(2)$ under the Schwinger
map~\cite{Schwinger1965,Arecchi1972,Sakurai2017}
\begin{equation}
  J_+ = a^\dagger b, \qquad J_- = a b^\dagger, \qquad
  J_z = \tfrac{1}{2}(a^\dagger a - b^\dagger b).
\end{equation}
The intrinsic code $\lcode\irrep{5},2,2\rcode$ is realized in the $N=4$ subspace via
the embedding $\emb: \ket{k} \mapsto \ket{k,4-k}$, giving codewords
\begin{align}
  \ket{\bar{0}} &= \tfrac{1}{\sqrt{2}}(\ket{0,4}+\ket{4,0}), \nonumber\\
  \ket{\bar{1}} &= \ket{2,2}.
\end{align}
This is precisely the Chuang--Leung--Yamamoto (CLY)
code~\cite{CLYcode}.
By the Schur bootstrap, it satisfies the KL condition for all errors transforming as
$\irrep{1}$ or $\irrep{3}$.

This covers two physically distinct error types.
For number-preserving (dephasing) errors: $\hat{n}$ transforms as $\irrep{1}$ and
$(J_+, J_-, J_z)$ transform as $\irrep{3}$, so all are detected.
Higher-order dephasing errors such as $J_+J_z - J_zJ_+$ are also detected whenever
they transform in $\irrep{1}$ or $\irrep{3}$.
For loss/gain errors: single-mode operators $\{a,b\}$ and $\{a^\dagger,b^\dagger\}$
transform as $\irrep{2}$, which does not appear in the decomposition~\eqref{eq:L(V)-su2-5},
so the code automatically satisfies the KL condition for all single-loss and single-gain
errors.
Mixed loss-gain operators such as $a^\dagger a$ transform as $\irrep{1} \oplus \irrep{3}$,
so these are detected as well.
Odd-order combinations of loss/gain operators contribute only even-dimensional irreps
of $\SU(2)$, none of which appear in $\L(V)$, so all are automatically detected.

\subsection{Two Spin-$1$ Qutrits: Depth Versus Conventional Distance}
\label{sec:example-qutrits}

The same intrinsic code also embeds into tensor products of higher-spin representations,
making the distinction between depth and conventional distance explicit.
For two spin-$1$ qutrits,
\begin{equation}
  \HH = (\irrep{3})^{\otimes 2} \;\cong\; \irrep{1} \oplus \irrep{3} \oplus \irrep{5}.
\end{equation}
The unique (up to phase) equivariant embedding $\emb: \irrep{5} \hookrightarrow \HH$
maps
\begin{align}
  \ket{0} &\mapsto \ket{00}, \nonumber\\
  \ket{1} &\mapsto \tfrac{1}{\sqrt{2}}(\ket{01}+\ket{10}), \nonumber\\
  \ket{2} &\mapsto \tfrac{1}{\sqrt{6}}(\ket{02}+2\ket{11}+\ket{20}), \nonumber\\
  \ket{3} &\mapsto \tfrac{1}{\sqrt{2}}(\ket{12}+\ket{21}), \nonumber\\
  \ket{4} &\mapsto \ket{22},
\end{align}
yielding codewords
\begin{align}
  \ket{\bar{0}} &= \tfrac{1}{\sqrt{2}}(\ket{00}+\ket{22}), \nonumber\\
  \ket{\bar{1}} &= \tfrac{1}{\sqrt{6}}(\ket{02}+2\ket{11}+\ket{20}).
\end{align}
The Schur bootstrap guarantees detection of all errors in the $\irrep{1}$ and $\irrep{3}$
sectors of $\L(\HH)$.
However, a single-site error on a spin-$1$ system decomposes as
\begin{equation}
  \L(\irrep{3}) \;=\; \irrep{1} \oplus \irrep{3} \oplus \irrep{5},
\end{equation}
so conventional distance $2$ would require detecting the entire single-site operator
space, including the $\irrep{5}$ component.
This code does not detect $\irrep{5}$ errors: it has depth $\mathsf{d} = 2$ but
conventional distance $1$.
This is the clearest illustration of the point that depth and weight-based distance
diverge outside the fundamental multi-qudit setting: the code suppresses all
leading-order ($\irrep{3}$) single-site errors while leaving subleading ($\irrep{5}$)
components undetected, a distinction that conventional distance cannot resolve.

\subsection{Trading sites for local dimensions}

The example above illustrates a general feature of the Schur bootstrap: the same
intrinsic code can be realized with varying numbers of sites and local dimensions,
with depth preserved throughout.
For the intrinsic code $\lcode\irrep{5},2,2\rcode$, this yields a hierarchy of
extrinsic realizations all sharing depth $\mathsf{d} = 2$:
\begin{itemize}
  \item one site: the spin-$2$ intrinsic code itself (local dimension $5$),
  \item two sites: the two-qutrit code above (local dimension $3$) or a
        qubit--ququart system $\irrep{2}\otimes\irrep{4}$ (local dimensions $2,4$),
  \item four sites: the $((4,2,2))$ permutation-invariant qubit code.
\end{itemize}
Moving along this hierarchy trades the number of physical sites for the local Hilbert
space dimension, while the symmetry-resolved error protection --- and hence
depth --- remains invariant by the Schur bootstrap.
The two-qutrit and qubit--ququart realizations have conventional distance $d = 1$
despite depth $\mathsf{d} = 2$, reflecting the divergence between weight-based and
representation-theoretic notions of protection outside the fundamental multi-qudit
setting.

\section{Example: A Distance-3 Intrinsic $\SU(2)$ Code with Clifford Symmetry}
\label{sec:example-su2-d3}

The previous example illustrated the Schur bootstrap at intrinsic distance
$\mathsf{d}=2$.
We now exhibit an intrinsic $\SU(2)$ code of distance $\mathsf{d}=3$ whose covariant
gate group is the single-qubit Clifford group, demonstrating that the intrinsic
framework simultaneously organizes higher-order error suppression and transversal gate
structure.

\subsection{An Intrinsic Code in $V = \irrep{14}$}
\label{sec:example2-intrinsic}

Let $G = \SU(2)$ and let $V = \irrep{14}$ be the spin-$\tfrac{13}{2}$ irrep.
Fix the weight basis $\{\ket{m}\}_{m=-13/2}^{13/2}$ and relabel as
$\ket{k} := \ket{m = \frac{13}{2}-k}$ for $k = 0,1,\ldots,13$.
Consider the two-dimensional subspace $\Cint \subset V$ spanned by
\begin{align}
  \ket{\bar{0}} &=
    \tfrac{\sqrt{5}}{112}(13\sqrt{2}+2\sqrt{11})\ket{0}
    + \tfrac{\sqrt{13}}{112}(10-3\sqrt{22})\ket{4} \nonumber\\
  &\quad - \tfrac{\sqrt{65}}{112}(6+\sqrt{22})\ket{8}
    + \tfrac{\sqrt{65}}{112}(\sqrt{2}-2\sqrt{11})\ket{12}, \nonumber \\
  \ket{\bar{1}} &=
    \tfrac{\sqrt{5}}{112}(13\sqrt{2}+2\sqrt{11})\ket{13}
    + \tfrac{\sqrt{13}}{112}(10-3\sqrt{22})\ket{9} \nonumber\\
  &\quad - \tfrac{\sqrt{65}}{112}(6+\sqrt{22})\ket{5}
    + \tfrac{\sqrt{65}}{112}(\sqrt{2}-2\sqrt{11})\ket{1}.
\end{align}
A direct computation shows $\Omega F \Omega = 0$ for all $F$ in the
$\irrep{3} \oplus \irrep{5}$ isotypic sectors of $\L(V)$, so $\Cint$ satisfies the
KL condition for all errors transforming in $\irrep{1}$, $\irrep{3}$, and $\irrep{5}$.
This establishes intrinsic distance $\mathsf{d} = 3$, and we denote the code
\begin{equation}
  \lcode\irrep{14}, 2, 3\rcode_{\SU(2)}.
\end{equation}
Compared to the first example, the KL conditions now eliminate not only the adjoint
sector $\irrep{3}$ but also the next isotypic sector $\irrep{5}$, corresponding to
protection against both first- and second-order errors in the adjoint hierarchy.

\subsection{Intrinsic Logical Symmetry}
\label{sec:example2-logical}

Certain elements of the $\SU(2)$ action on $V = \irrep{14}$ preserve $\Cint$ and
enact nontrivial logical operations.
Working in $\SU(2)$, let $\mathsf{X}, \mathsf{Z}, \mathsf{H}, \mathsf{S}$ denote
the standard single-qubit Pauli, Hadamard, and phase gates viewed as elements of
$\SU(2)$.
Their $14$-dimensional representations act on $\Cint$ as follows:
\begin{align*}
    \rho_{14}(\mathsf{X}) &\mapsto\; \bar{\mathsf{X}}, \\
  \rho_{14}(\mathsf{Z}) &\mapsto\; \bar{\mathsf{Z}}, \\
  \rho_{14}(\mathsf{S}^{3\dagger}) &\mapsto\; \bar{\mathsf{S}}, \\
  \rho_{14}(\mathsf{H}^\dagger) &\mapsto\; \bar{\mathsf{H}}
\end{align*}
Together these generate the single-qubit Clifford group acting
faithfully on $\Cint$, so $\lcode\irrep{14},2,3\rcode$ is Clifford covariant.
By the intrinsic Eastin--Knill theorem (Lemma~\ref{lem:IntrinsicEK}), this is
consistent with distance $\mathsf{d} = 3$: since the Clifford group is finite, the
discreteness requirement is satisfied.

\subsection{Thirteen Qubits: A Permutation-Invariant $((13,2,3))$ Code}
\label{sec:example2-13qubit}

Since $\irrep{14}$ has half-integer spin, it appears in $(\irrep{2})^{\otimes n}$ only
for odd $n$.
The smallest realization occurs at $n = 13$, where $\irrep{14}$ appears with
multiplicity one in the permutation-invariant subspace, yielding a unique
$((13,2,3))$ permutation-invariant qubit code.

By the Schur bootstrap (Lemma~\ref{lem:SchurBootstrap}), this code detects all error
components transforming in $\irrep{1}$, $\irrep{3}$, and $\irrep{5}$.
In the multi-qubit setting every weight-$1$ error transforms in $\irrep{3}$ and every
weight-$2$ error decomposes into irreps $ \irrep{1}, \irrep{3}, \irrep{5} $, so intrinsic distance
$\mathsf{d} = 3$ coincides with conventional distance $d = 3$.

Because the $\SU(2)$ action on $(\CC^2)^{\otimes 13}$ is $g^{\otimes 13}$, the
intrinsic logical symmetry lifts directly to a transversal implementation.
Letting $X, Z, H, S$ denote the standard single-qubit gates, the physical operators
\begin{equation}
  X^{\otimes 13}, \quad Z^{\otimes 13}, \quad H^{\otimes 13}, \quad S^{\otimes 13}
\end{equation}
preserve the codespace and implement the logical gates
\begin{equation}
  \bar{X}, \quad \bar{Z}, \quad -\bar{H}, \quad \bar{S},
\end{equation}
respectively.
The entire single-qubit Clifford group is therefore implemented transversally on this
$((13,2,3))$ code.

\subsection{Moduli Spaces of Higher Qubit Realizations}
\label{sec:example2-moduli}

For odd $n \geq 15$, the irrep $\irrep{14}$ appears in $(\CC^2)^{\otimes n}$ with
multiplicity
\begin{equation}
  \mu \;=\; \frac{28}{n+15}\binom{n}{\frac{n-13}{2}} \;\sim\; \mathcal{O}(2^n),
  \label{eq:irep14-multiplicity}
\end{equation}
giving a $\CC P^{\mu-1}$ family of $((n,2,3))$ codes.
For example, $n = 15$ gives multiplicity $14$ and a $\CC P^{13}$ family, while
$n = 17$ gives multiplicity $119$ and a $\CC P^{118}$ family.
By Corollary~\ref{cor:depth-preserved} and the equivariance of the embedding,
every member of each family inherits both distance $d = 3$ and a transversal
Clifford gate set.

\subsection{Further Realizations}
\label{sec:example2-further}

The same intrinsic code admits realizations beyond qubit tensor powers, all
inheriting depth $\mathsf{d} = 3$ and transversal Clifford symmetry by the Schur
bootstrap.

For homogeneous higher-spin systems, $\irrep{14}$ appears in
$(\irrep{4})^{\otimes 5}$ with multiplicity $2$ (giving a $\CC P^1$ family) and in
$(\irrep{6})^{\otimes 3}$ with multiplicity $4$ (giving a $\CC P^3$ family).

For heterogeneous systems, $\irrep{14}$ appears (with multiplicity one) in a variety
of tensor products, including $\irrep{7} \otimes \irrep{8}$ (two sites),
$\irrep{4} \otimes \irrep{6} \otimes \irrep{6}$ (three sites),
$\irrep{4}^{\otimes 3} \otimes \irrep{5}$ (four sites), and
$\irrep{2} \otimes \irrep{3}^{\otimes 6}$ (seven sites).
The last realization is notable: it has conventional qudit distance $d = 2$ in
addition to intrinsic depth $\mathsf{d} = 3$, since the $\irrep{5}$ component of a
single spin-$1$ site error is detectable by the depth condition while the $\irrep{3}$
component is fully corrected.

A structural contrast with the first example is worth noting.
The irrep $\irrep{5}$ is odd-dimensional and therefore occurs in rotational Hilbert
spaces decomposing under $\SO(3)$, giving rise to the linear rotor, rigid rotor, and
molecular realizations explored in the supplemental materia.
The irrep $\irrep{14}$ is even-dimensional and does not appear in purely rotational
$\SO(3)$ decompositions; realizations in molecular rotation require a double group
incorporating electronic or nuclear spin degrees of freedom.

Together, the two examples demonstrate that intrinsic quantum codes naturally
organize higher-order error protection, covariant gate structure, and continuous
moduli spaces of extrinsic realizations within a single representation-theoretic
framework.

\section{Example: A Distance-2 Intrinsic $\SU(3)$ Code with $A_6$ Logical Symmetry}

We illustrate the intrinsic framework with a concrete example for $G=\SU(3)$.
Let $V$ be the irreducible representation
\[
V = \dynkin{2,2} \cong \irrep{27}. \numberthis
\]
We construct an intrinsic code $\Cint \subset V$ of dimension $\dim \Cint = 5$.

Let $3.A_6 \subset \SU(3)$ be the finite subgroup of order $1080$ appearing in
\cite{us3,us4} (also known as $\Sigma(360\phi)$). The restriction of $V$ to $3.A_6$
decomposes as
\begin{equation}\label{eq:27-branching}
\irrep{27}\!\downarrow_{3.A_6}
=
\chi_6 \oplus \chi_7 \oplus \chi_{11} \oplus \chi_{12},
\end{equation}
where $\chi_6$ and $\chi_7$ are the two inequivalent $5$-dimensional irreducible
representations (descending to the two degree-$5$ irreps of $A_6$), and
$\chi_{11},\chi_{12}$ are higher-dimensional constituents (see \cite{us3} for notation). 

We define the intrinsic code $\Cint$ to be the $\chi_6$-isotypic subspace:
\[
\Cint \;\cong\; \chi_6 \subset V. \numberthis
\]
Thus $\dim \Cint = 5$. An explicit orthonormal basis $\{\ket{\overline{0}},\ldots,\ket{\overline{4}}\}$
for $\Cint$ is given in \cref{app:explicitcodewords} of the Appendix.

The setwise stabilizer of $\Cint$ in $\SU(3)$ is
\[
\G_{\mathrm{inv}} = 3.A_6, \numberthis
\]
and the pointwise stabilizer is the central subgroup
\[
\G_0 = \langle \zeta_3 \mathbf{1} \rangle \cong \ZZ_3. \numberthis
\]
Therefore the induced logical symmetry group is
\[
\GG = \G_{\mathrm{inv}}/\G_0 \cong A_6. \numberthis
\]

The code $\Cint$ has distance $d=2$ which can be checked using the explicit codewords in the appendix. More rigorously, this follows directly from the twisted unitary $1$-group construction of
\cite{us3}. In the intrinsic formulation, this corresponds to the Knill--Laflamme condition
holding on the adjoint sector, i.e., intrinsic depth $\mathsf d \ge 2$.

Using our notation we would label this code as
\[
  \lcode\irrep{27}, 5, 2\rcode_{\SU(3)}. \numberthis
\]

\subsection{Multiple realizations}

The intrinsic representation $V \cong \irrep{27}$ admits multiple
$\SU(3)$-equivariant embeddings into multi-qutrit Hilbert spaces.
In particular, there is a unique copy of $V$ in both $6$ qutrits (${\irrep{3}}^{\otimes 6}$) using the Young Embedding and $4$ qutrits ($\irrep{3} \otimes \irrep{3} \otimes \irrepbar{3} \otimes \irrepbar{3}$) using the Canonical Embedding (see Appendix~\ref{sec:embeddings} for more detail).
Consequently, the intrinsic code $\Cint$ yields both a $6$-qutrit distance-$2$ code inside $(\CC^3)^{\otimes 6}$, and a $4$-qutrit distance-$2$ code inside $\CC^3 \otimes \CC^3 \otimes \overline{\CC^3} \otimes \overline{\CC^3}$.

In these realizations, the $\SU(3)$ action implements logical gates
in a simple tensor-product form.
For the $6$-qutrit realization, the physical action $g^{\otimes 6}$ acts on the codespace as the logical representation $\chi_6(g)$.
Similarly, in the $4$-qutrit realization,
\[
g \otimes g \otimes \bar{g} \otimes \bar{g}
\]
implements the same logical action $\chi_6(g)$ on $\Cint$.

It is remarkable to note that decomposition \eqref{eq:27-branching} also contains a second
$5$-dimensional irreducible component $\chi_7$.
This yields a second intrinsic code $\Cint' \subset V$ of dimension $5$,
which likewise has distance $d=2$ and logical symmetry group $A_6$.
The representations $\chi_6$ and $\chi_7$ are exchanged by an outer
automorphism of $A_6$, so the corresponding codes are equivalent as
abstract $A_6$-modules up to outer automorphism, but are distinct
subspaces of $V$. The companion $\chi_7$ construction also produces codes in both
the $4$- and $6$-qutrit realizations, with the same distance and
logical symmetry group $A_6$, but corresponding to the second
degree-$5$ irreducible representation of $A_6$.

\subsection{Relation to previous constructions}

This example recovers, in intrinsic form, the $3.A_6$ construction of
\cite{us4}.
In that work, $3.A_6 \subset \SU(3)$ appears as a twisted unitary
$1$-group producing distance-$2$, $ 6$ qutrit codes with logical symmetry $A_6$.
Here, the same structure arises from a single intrinsic
$\SU(3)$ representation $V=\irrep{27}$, whose restriction to $3.A_6$
contains the relevant $5$-dimensional constituents.
The intrinsic formulation makes clear that these codes are not isolated
constructions, but rather arise from a common representation-theoretic
mechanism and admit multiple inequivalent physical realizations.

\medskip

This example illustrates that a single intrinsic representation can encode
multiple inequivalent logical sectors and admit multiple physically distinct
realizations, all governed by the same underlying symmetry.

\section{Conclusion}

We have introduced intrinsic quantum codes as subspaces $\Cint \subset V$ of a
unitary representation $V$ of a symmetry group $G$, with error protection encoded
representation-theoretically through the isotypic decomposition of $\L(V)$.
Our main structural result, the Schur bootstrap, shows that intrinsic
Knill--Laflamme conditions are preserved under every $G$-equivariant isometric
embedding of $V$ into a larger Hilbert space.
Thus, a single verification in the intrinsic representation certifies error
protection across an entire family of physical realizations.

For $G=\SU(q)$, the adjoint-order hierarchy induces a notion of intrinsic depth
that recovers conventional code distance for fundamental multi-qudit systems and
refines it for higher or heterogeneous local representations.
In this sense, both distance and depth become invariants of the representation,
transported automatically to all equivariant realizations.
This places earlier symmetry-based constructions~\cite{us2} in a unified
representation-theoretic framework independent of any particular embedding.

We also proved an intrinsic Eastin--Knill theorem: any intrinsic code of depth
$\mathsf d \ge 2$ admits only a discrete logical symmetry group, and for compact
$G$ this group is finite.
Combined with the Schur bootstrap, this yields transversality obstructions for all
extrinsic realizations from a single representation-theoretic argument, showing
that the obstruction to continuous covariant logical gates is already present at
the intrinsic level.

The examples illustrate the range of this framework.
The minimal code $\lcode\irrep{5},2,2\rcode_{\SU(2)}$ underlies the four-qubit
permutation-invariant code, the CLY bosonic code, a $\CC P^4$ family of six-qubit
codes, two-qutrit and qubit-ququart realizations, and molecular codes.
The code $\lcode\irrep{14},2,3\rcode_{\SU(2)}$ yields exponentially large families
of permutation-invariant qubit codes with transversal single-qubit Clifford gates,
as well as heterogeneous multi-qudit realizations.
The $\SU(3)$ example in the irrep $\irrep{27}$ shows that a single intrinsic
distance-$2$ code can recover earlier $3.A_6$ constructions and produce both
$6$-qutrit and $4$-qutrit realizations with logical symmetry group $A_6$.
In each case, properties certified once in a small representation space propagate
uniformly to all equivariant realizations.

Several directions remain open.
It would be natural to extend the framework to codes with multiple symmetries and
to approximate quantum error correction in the intrinsic setting.
The moduli spaces of extrinsic realizations, parameterized by $\CC P^{m-1}$ when a
target irrep appears with multiplicity $m$, also raise natural optimization
questions, for example for gate universality or robustness to non-symmetric noise.
More broadly, the representation-theoretic viewpoint developed here may connect to
other symmetry-based structures in quantum information, including covariant codes
for continuous symmetries~\cite{CovariantCatPRL}, holographic quantum error
correction, and the classification of fault-tolerant gate sets beyond the
Eastin--Knill obstruction.

\section*{Acknowledgments}

We thank Victor Albert and Anthony Leverrier for helpful conversations on bosonic codes, molecular codes, codes made from mixed irreps, and other types of covariant quantum codes outside of the multiqudit framework.

\bibliographystyle{IEEEtran}
\bibliography{biblio.bib}

\begin{thebibliography}{10}
\providecommand{\url}[1]{#1}
\csname url@samestyle\endcsname
\providecommand{\newblock}{\relax}
\providecommand{\bibinfo}[2]{#2}
\providecommand{\BIBentrySTDinterwordspacing}{\spaceskip=0pt\relax}
\providecommand{\BIBentryALTinterwordstretchfactor}{4}
\providecommand{\BIBentryALTinterwordspacing}{\spaceskip=\fontdimen2\font plus
\BIBentryALTinterwordstretchfactor\fontdimen3\font minus \fontdimen4\font\relax}
\providecommand{\BIBforeignlanguage}[2]{{%
\expandafter\ifx\csname l@#1\endcsname\relax
\typeout{** WARNING: IEEEtran.bst: No hyphenation pattern has been}%
\typeout{** loaded for the language `#1'. Using the pattern for}%
\typeout{** the default language instead.}%
\else
\language=\csname l@#1\endcsname
\fi
#2}}
\providecommand{\BIBdecl}{\relax}
\BIBdecl

\bibitem{gauss1828disquisitiones}
C.~F. Gauss, \emph{Disquisitiones Generales Circa Superficies Curvas}.\hskip 1em plus 0.5em minus 0.4em\relax G{\"o}ttingen: Dieterich, 1828, g{\"o}ttingen: Dieterich, 1828; translated in {\it General Investigations of Curved Surfaces}, Princeton Univ. Press (1902).

\bibitem{gross1}
\BIBentryALTinterwordspacing
J.~A. Gross, ``Designing codes around interactions: The case of a spin,'' \emph{Phys. Rev. Lett.}, vol. 127, p. 010504, Jul 2021. [Online]. Available: \url{https://link.aps.org/doi/10.1103/PhysRevLett.127.010504}
\BIBentrySTDinterwordspacing

\bibitem{CovariantCatPRL}
\BIBentryALTinterwordspacing
A.~Denys and A.~Leverrier, ``Quantum error-correcting codes with a covariant encoding,'' \emph{Phys. Rev. Lett.}, vol. 133, p. 240603, Dec 2024. [Online]. Available: \url{https://link.aps.org/doi/10.1103/PhysRevLett.133.240603}
\BIBentrySTDinterwordspacing

\bibitem{TessellationCodesPRL}
\BIBentryALTinterwordspacing
Y.~Wang, Y.~Xu, and Z.-W. Liu, ``Tessellation codes: Encoded quantum gates by geometric rotation,'' \emph{Phys. Rev. Lett.}, vol. 135, p. 140602, Oct 2025. [Online]. Available: \url{https://link.aps.org/doi/10.1103/tljb-f7tt}
\BIBentrySTDinterwordspacing

\bibitem{us1}
\BIBentryALTinterwordspacing
E.~Kubischta and I.~Teixeira, ``Family of quantum codes with exotic transversal gates,'' \emph{Phys. Rev. Lett.}, vol. 131, p. 240601, Dec 2023. [Online]. Available: \url{https://link.aps.org/doi/10.1103/PhysRevLett.131.240601}
\BIBentrySTDinterwordspacing

\bibitem{us3}
\BIBentryALTinterwordspacing
------, ``Quantum codes from twisted unitary $t$-groups,'' \emph{Phys. Rev. Lett.}, vol. 133, p. 030602, Jul 2024. [Online]. Available: \url{https://link.aps.org/doi/10.1103/PhysRevLett.133.030602}
\BIBentrySTDinterwordspacing

\bibitem{RuskaiPRL}
\BIBentryALTinterwordspacing
M.~B. Ruskai, ``Pauli exchange errors in quantum computation,'' \emph{Phys. Rev. Lett.}, vol.~85, pp. 194--197, Jul 2000. [Online]. Available: \url{https://link.aps.org/doi/10.1103/PhysRevLett.85.194}
\BIBentrySTDinterwordspacing

\bibitem{2004permutation}
\BIBentryALTinterwordspacing
H.~Pollatsek and M.~B. Ruskai, ``Permutationally invariant codes for quantum error correction,'' \emph{Linear Algebra and its Applications}, vol. 392, pp. 255--288, 2004. [Online]. Available: \url{https://www.sciencedirect.com/science/article/pii/S0024379504002903}
\BIBentrySTDinterwordspacing

\bibitem{ouyangPI}
\BIBentryALTinterwordspacing
Y.~Ouyang, ``Permutation-invariant quantum codes,'' \emph{Physical Review A}, vol.~90, no.~6, Dec. 2014. [Online]. Available: \url{http://dx.doi.org/10.1103/PhysRevA.90.062317}
\BIBentrySTDinterwordspacing

\bibitem{Hagiwara2020FourQubitDeletion}
\BIBentryALTinterwordspacing
M.~Hagiwara and A.~Nakayama, ``A four-qubits code that is a quantum deletion error-correcting code with the optimal length,'' in \emph{2020 International Symposium on Information Theory (ISIT)}, 2020, pp. 917--921. [Online]. Available: \url{https://ieeexplore.ieee.org/document/9174339}
\BIBentrySTDinterwordspacing

\bibitem{Nakayama2020SingleDeletion}
\BIBentryALTinterwordspacing
A.~Nakayama and M.~Hagiwara, ``Single quantum deletion error-correcting codes,'' in \emph{2021 International Symposium on Information Theory (ISIT)}, 2021, pp. 1500--1504. [Online]. Available: \url{https://ieeexplore.ieee.org/document/9366129}
\BIBentrySTDinterwordspacing

\bibitem{Ouyang2021PermutationInvariant}
\BIBentryALTinterwordspacing
Y.~Ouyang, ``Permutation-invariant quantum coding for quantum deletion channels,'' in \emph{2021 IEEE International Symposium on Information Theory (ISIT)}, 2021, pp. 1499--1504. [Online]. Available: \url{https://ieeexplore.ieee.org/document/9517952}
\BIBentrySTDinterwordspacing

\bibitem{422codePRL}
\BIBentryALTinterwordspacing
S.~Bravyi, D.~Lee, Z.~Li, and B.~Yoshida, ``How much entanglement is needed for quantum error correction?'' \emph{Phys. Rev. Lett.}, vol. 134, p. 210602, May 2025. [Online]. Available: \url{https://link.aps.org/doi/10.1103/PhysRevLett.134.210602}
\BIBentrySTDinterwordspacing

\bibitem{CLYcode}
\BIBentryALTinterwordspacing
I.~L. Chuang, D.~W. Leung, and Y.~Yamamoto, ``Bosonic quantum codes for amplitude damping,'' \emph{Phys. Rev. A}, vol.~56, pp. 1114--1125, Aug 1997. [Online]. Available: \url{https://link.aps.org/doi/10.1103/PhysRevA.56.1114}
\BIBentrySTDinterwordspacing

\bibitem{fultonharris}
W.~Fulton and J.~Harris, \emph{Representation theory}, 1st~ed., ser. Graduate texts in mathematics.\hskip 1em plus 0.5em minus 0.4em\relax New York, NY: Springer, Jul. 1999.

\bibitem{Georgi}
H.~Georgi, \emph{{Lie algebras in particle physics}}, 2nd~ed.\hskip 1em plus 0.5em minus 0.4em\relax Reading, MA: Perseus Books, 1999, vol.~54.

\bibitem{supp}
``See {S}upplemental {M}aterial which includes references \cite{GroupMath, mathematica, rainsWE, jacobsonsphere, monopoleharmonics, YaleFan, gross2, Bunker_Jensen_1998, PI2, PI3, PIqudit1, ACP, victornew, 2004permutation, RuskaiPRL, us2}.''

\bibitem{Schwinger1965}
J.~Schwinger, ``On angular momentum,'' in \emph{Quantum Theory of Angular Momentum}, L.~C. Biedenharn and H.~van Dam, Eds.\hskip 1em plus 0.5em minus 0.4em\relax Academic Press, 1965, pp. 229--279.

\bibitem{Arecchi1972}
F.~T. Arecchi, E.~Courtens, R.~Gilmore, and H.~Thomas, ``Atomic coherent states in quantum optics,'' \emph{Physical Review A}, vol.~6, no.~6, pp. 2211--2237, 1972.

\bibitem{Sakurai2017}
J.~J. Sakurai and J.~Napolitano, \emph{Modern Quantum Mechanics}, 2nd~ed.\hskip 1em plus 0.5em minus 0.4em\relax Pearson, 2017.

\bibitem{us4}
\BIBentryALTinterwordspacing
E.~Kubischta and I.~Teixeira, ``Quantum codes and irreducible products of characters,'' \emph{Designs, Codes and Cryptography}, Apr. 2025. [Online]. Available: \url{http://dx.doi.org/10.1007/s10623-025-01599-8}
\BIBentrySTDinterwordspacing

\bibitem{us2}
------, ``Permutation-invariant quantum codes with transversal generalized phase gates,'' \emph{IEEE Transactions on Information Theory}, vol.~71, no.~1, pp. 485--498, 2025.

\bibitem{GroupMath}
\BIBentryALTinterwordspacing
R.~M. Fonseca, ``Groupmath: A mathematica package for group theory calculations,'' \emph{Computer Physics Communications}, vol. 267, p. 108085, 2021. [Online]. Available: \url{https://www.sciencedirect.com/science/article/pii/S0010465521001971}
\BIBentrySTDinterwordspacing

\bibitem{mathematica}
\BIBentryALTinterwordspacing
{Wolfram Research, Inc.}, ``Mathematica 14.'' [Online]. Available: \url{https://www.wolfram.com}
\BIBentrySTDinterwordspacing

\bibitem{rainsWE}
E.~Rains, ``Quantum weight enumerators,'' \emph{IEEE Transactions on Information Theory}, vol.~44, no.~4, pp. 1388--1394, 1998.

\bibitem{jacobsonsphere}
\BIBentryALTinterwordspacing
R.~Andrade~e Silva and T.~Jacobson, ``Particle on the sphere: group-theoretic quantization in the presence of a magnetic monopole,'' \emph{Journal of Physics A: Mathematical and Theoretical}, vol.~54, no.~23, p. 235303, May 2021. [Online]. Available: \url{http://dx.doi.org/10.1088/1751-8121/abf961}
\BIBentrySTDinterwordspacing

\bibitem{monopoleharmonics}
T.~Dray, ``The relationship between monopole harmonics and spin-weighted spherical harmonics,'' \emph{J. Math. Phys.}, vol.~26, no.~5, pp. 1030--1033, May 1985.

\bibitem{YaleFan}
\BIBentryALTinterwordspacing
Y.~Fan, W.~Fischler, and E.~Kubischta, ``Quantum error correction in the lowest landau level,'' \emph{Physical Review A}, vol. 107, no.~3, Mar. 2023. [Online]. Available: \url{http://dx.doi.org/10.1103/PhysRevA.107.032411}
\BIBentrySTDinterwordspacing

\bibitem{gross2}
S.~Omanakuttan and J.~A. Gross, ``Multispin clifford codes for angular momentum errors in spin systems,'' 2023.

\bibitem{Bunker_Jensen_1998}
P.~R. Bunker and P.~Jensen, \emph{Molecular symmetry and spectroscopy 2nd ed.}\hskip 1em plus 0.5em minus 0.4em\relax NRC Research Press, 1998.

\bibitem{PI2}
\BIBentryALTinterwordspacing
Y.~Ouyang, ``Permutation-invariant quantum codes,'' \emph{Physical Review A}, vol.~90, no.~6, Dec. 2014. [Online]. Available: \url{http://dx.doi.org/10.1103/PhysRevA.90.062317}
\BIBentrySTDinterwordspacing

\bibitem{PI3}
\BIBentryALTinterwordspacing
A.~Aydin, M.~A. Alekseyev, and A.~Barg, ``A family of permutationally invariant quantum codes,'' \emph{Quantum}, vol.~8, p. 1321, Apr. 2024. [Online]. Available: \url{http://dx.doi.org/10.22331/q-2024-04-30-1321}
\BIBentrySTDinterwordspacing

\bibitem{PIqudit1}
\BIBentryALTinterwordspacing
Y.~Ouyang, ``Permutation-invariant qudit codes from polynomials,'' \emph{Linear Algebra and its Applications}, vol. 532, p. 43–59, Nov. 2017. [Online]. Available: \url{http://dx.doi.org/10.1016/j.laa.2017.06.031}
\BIBentrySTDinterwordspacing

\bibitem{ACP}
\BIBentryALTinterwordspacing
V.~V. Albert, J.~P. Covey, and J.~Preskill, ``Robust encoding of a qubit in a molecule,'' \emph{Physical Review X}, vol.~10, no.~3, Sep. 2020. [Online]. Available: \url{http://dx.doi.org/10.1103/PhysRevX.10.031050}
\BIBentrySTDinterwordspacing

\bibitem{victornew}
\BIBentryALTinterwordspacing
A.~Aydin, V.~V. Albert, and A.~Barg, ``Quantum error correction beyond {SU(2)}: Spin, bosonic, and permutation-invariant codes from convex geometry,'' \emph{PRX Quantum}, vol.~7, no.~1, Feb. 2026. [Online]. Available: \url{http://dx.doi.org/10.1103/kx3b-4nrp}
\BIBentrySTDinterwordspacing

\bibitem{Shubham}
\BIBentryALTinterwordspacing
S.~P. Jain, E.~R. Hudson, W.~C. Campbell, and V.~V. Albert, ``Absorption-emission codes for atomic and molecular quantum information platforms,'' \emph{Phys. Rev. Lett.}, vol. 133, p. 260601, Dec 2024. [Online]. Available: \url{https://link.aps.org/doi/10.1103/PhysRevLett.133.260601}
\BIBentrySTDinterwordspacing

\bibitem{AEbarg}
\BIBentryALTinterwordspacing
A.~Aydin and A.~Barg, ``Class of codes correcting absorptions and emissions,'' \emph{Physical Review A}, vol. 111, no.~2, Feb. 2025. [Online]. Available: \url{http://dx.doi.org/10.1103/PhysRevA.111.022415}
\BIBentrySTDinterwordspacing

\end{thebibliography}

\appendices

\makeatletter
\renewcommand{\theequation}{A\arabic{equation}}
\renewcommand{\thetable}{A\arabic{table}}
\renewcommand{\thefigure}{A\arabic{figure}}
\renewcommand{\thelemma}{A\arabic{lemma}}
\renewcommand{\thetheorem}{A\arabic{theorem}}
\setcounter{table}{0}
\setcounter{figure}{0}
\setcounter{lemma}{0}
\setcounter{theorem}{0}
\setcounter{equation}{0}

\section{Minimal Embeddings of $\SU(q)$ intrinsic irrep codes}
\label{sec:embeddings}

Given an irrep $V$ of $\SU(q)$, labeled by Dynkin indices $(p_1,\ldots,p_{q-1})$,
we describe three standard $\SU(q)$-equivariant isometric embeddings of $V$ into a
multi-site Hilbert space $\HH = \bigotimes_i \HH_i$.
Each embedding has distinct physical relevance: the canonical embedding minimizes the
number of sites at the cost of heterogeneous local spaces; the natural multi-qudit
embedding uses only fundamental and anti-fundamental qudits; and the Young embedding
uses only fundamental qudits at the cost of more sites.
Let $\irrep{F} = \CC^q$ and $\irrepbar{F} = \overline{\CC^q}$ denote the fundamental
and anti-fundamental irreps of $\SU(q)$, respectively.

\subsection*{Canonical Embedding. } The minimal canonical Hilbert space $\HH$ that contains the irrep $V$ with multiplicity $1$ is 
\[
    \HH = \bigotimes_{k=1}^{q-1} \qty(\wedge^k(\irrep{F}))^{\otimes p_k}, \numberthis
\]
where $\wedge^k(\irrep{F})$ is the $k$-fold wedge product of the fundamental irrep $\irrep{F}$. Note that $\wedge^1(\irrep{F}) = \irrep{F}$ and $\wedge^{q-1}(\irrep{F}) = \irrepbar{F}$ and $\wedge^k(\irrep{F}) = \wedge^{q-k}(\irrepbar{F})$. The natural $G = \SU(q)$ action for $g \in G$ is by
\[
     \bigotimes_{k=1}^{q-1} \qty(\wedge^k(g))^{\otimes p_k}. \numberthis
\]
Again $\wedge^1(g) = g$ and $\wedge^{q-1}(g) = g^*$ and $\wedge^k(g) = \wedge^{q-k}(g^*)$.

For example the irrep $V = \dynkin{2,1} = \irrep{15}$ of $\SU(3)$ can be embedded into $3$ sites using $\HH = \irrep{3} \otimes \irrep{3} \otimes \irrepbar{3}$ with action $g \otimes g \otimes g^*$. As another example, the irrep $V = \dynkin{2,1,1} = \irrepbar{140}$ of $\SU(4)$ can be embedded into 4 sites using
\[
\HH = \irrep{4} \otimes \irrep{4} \otimes \irrep{6} \otimes \irrepbar{4}, \numberthis
\]
where we have used the fact that $\irrep{4} \wedge \irrep{4} = \irrep{6}$ in $\SU(4)$. The natural action is by $g \otimes g \otimes (g \wedge g) \otimes g^*$.  

In general, this embedding has $n = \sum_{k=1}^{q-1} p_k$ sites and if $q > 3$ and any of the $p_2, \cdots,p_{q-2}$ are non-zero then this embedding will be heterogeneous (as the last example shows).

\subsection*{Natural Multi-qudit Embedding. } To get a natural strictly multi-qudit embedding, we can simply consider every $\wedge^k(\irrep{F})$ factor as a subspace of a tensor product of fundamental irreps $\irrep{F}$ and anti-fundamental irreps $\irrepbar{F}$. The natural $g \in \SU(q)$ action is by tensor products of $g$ (whenever there is an $\irrep{F}$) and $g^*$ (whenever there is an $\irrepbar{F}$).

In general $\wedge^k \irrep{F} \hookrightarrow {\irrep{F}}^{\otimes k}$ and $\wedge^k \irrep{F} \simeq \wedge^{q-k} \irrepbar{F} \hookrightarrow {\irrepbar{F}}^{\otimes (q-k)}$. Thus when $k$ is small ${\irrep{F}}^{\otimes k}$ is the smaller embedding and when $k$ is large then ${\irrepbar{F}}^{\otimes (q-k)}$ is the smaller embedding. In particular, when $q$ is odd we have
\[
    \wedge^k \irrep{F} \hookrightarrow \begin{cases}
        {\irrep{F}}^{\otimes k} & 1 \leq k \leq \tfrac{q-1}{2} \\
        {\irrepbar{F}}^{\otimes (q-k) } & \tfrac{q+1}{2} \leq k \leq q-1
    \end{cases} \numberthis
\]
When $q$ is even we have
\[
    \wedge^k \irrep{F} \hookrightarrow \begin{cases}
        {\irrep{F}}^{\otimes k} & 1 \leq k \leq \tfrac{q}{2}-1 \\
        {\irrep{F}}^{\otimes k} \text{ or } {\irrepbar{F}}^{\otimes (q-k)} & k = \tfrac{q}{2} \\
        {\irrepbar{F}}^{\otimes (q-k) } & \tfrac{q}{2}+1 \leq k \leq q-1
    \end{cases}. \numberthis
\]
So the minimal number of qudits we can embed irrep $V = (p_1, p_2, \cdots, p_{q-1})$ of $\SU(q)$ into is
\[
    n = \begin{cases}
        \sum_{k=1}^{(q-1)/2} k p_k + \sum_{k = (q+1)/2}^{q-1} (q-k)p_k & q \text{ odd} \\
        \sum_{k=1}^{\tfrac{q}{2} -1 } k p_k + \tfrac{q}{2} p_{q/2} + \sum_{k = \tfrac{q}{2} + 1 }^{q-1} (q-k)p_k & q \text{ even}
    \end{cases} \numberthis
\]
This can be written more compactly as 
\[
    n = \sum_{k=1}^{q-1} \min(k, q-k) p_k. \numberthis
\]
Note that this is no longer a unique embedding.

In summary, we need more sites for this embedding than before but now the embedding is multi-qudit instead of heterogeneous. Note that for $\SU(2)$ and $\SU(3)$ or for $\SU(q)$ where $p_2 = \cdots = p_{q-2} = 0$ this embedding is equivalent to the canonical embedding.

\subsection*{Young Embedding. } Lastly, consider the Young embedding. This is also multi-qudit but it only uses fundamental irreps $\irrep{F}$ and no anti-fundamental irreps $\irrepbar{F}$.

This picture is in some ways the easiest to understand because you just match the irrep $(p_1,p_2, \cdots, p_{q-1})$  to a Young diagram with
\[
n = \sum_{k=1}^{q-1} k p_k \numberthis
\]
boxes with row lengths $\ell_r = \sum_{i=r}^{q-1} p_i$. The action $g^{\otimes n}$ on $\HH = \irrep{F}^{\otimes n}$ acts irreducibly on an invariant
subspace $\emb(V)$ obtained by symmetrizing within each row and antisymmetrizing
within each column of the Young diagram.
This embedding is also non-unique.
It is the natural choice when the physical platform cannot naturally realize the
anti-fundamental representation $\irrepbar{F}$; for qubit systems ($q=2$) it
coincides with the permutation-invariant embedding used throughout the main text.

\begin{table}[h]
\centering
\begin{tabular}{lccc}
\hline\hline
Embedding & Local spaces & Sites $n$  \\
\hline
Canonical & $\wedge^k(\irrep{F})$ (heterogeneous) &
  $\sum_k p_k$  \\
Natural multi-qudit & $\irrep{F}$, $\irrepbar{F}$ &
  $\sum_k \min(k,q{-}k)\,p_k$  \\
Young & $\irrep{F}$ only &
  $\sum_k k\,p_k$  \\
\hline\hline
\end{tabular}
\caption{Comparison of the three standard minimal embeddings of an $\SU(q)$ irrep
$(p_1,\ldots,p_{q-1})$ into a multi-site Hilbert space.}
\label{tab:embeddings}
\end{table}

\section{Parameterizing Six-Qubit Realizations of $\lcode\irrep{5},2,2\rcode$}
\label{app:6qubit}

We give an explicit parameterization of the $\CC P^4$ family of $((6,2,2))$ codes
arising from the intrinsic code $\lcode\irrep{5},2,2\rcode$ via the Schur bootstrap.
The basis for the moduli space was computed using the GroupMath
package~\cite{GroupMath} in \textsc{Mathematica}~\cite{mathematica}.

For $\HH = (\CC^2)^{\otimes 6}$, the logical state $\ket{\bar{0}}$ takes the form
\begin{align}
  \ket{\bar{0}} \;=\;
  &\; c_1\bigl(\ket{100000}-\ket{011111}\bigr)
  + c_2\bigl(\ket{010000}-\ket{101111}\bigr) \nonumber\\
  &+ c_3\bigl(\ket{001000}-\ket{110111}\bigr)
  + c_4\bigl(\ket{000100}-\ket{111011}\bigr) \nonumber\\
  &+ c_5\bigl(\ket{000010}-\ket{111101}\bigr)
  + c_6\bigl(\ket{000001}-\ket{111110}\bigr),
  \label{eq:6qubit-log0}
\end{align}
where the coefficients $c_k$ are linear functions of parameters
$z = (z_1,\ldots,z_5) \in \CC P^4$:
\begin{align}
  c_1 &= \tfrac{1}{2}\sqrt{\tfrac{5}{3}}\,z_1, \nonumber\\
  c_2 &= -\tfrac{z_1}{2\sqrt{15}} + \sqrt{\tfrac{2}{5}}\,z_2, \nonumber\\
  c_3 &= \tfrac{1}{60}\!\left(-2\sqrt{15}\,z_1 - 3\sqrt{10}\,z_2
        + 15\sqrt{6}\,z_3\right), \nonumber\\
  c_4 &= \tfrac{1}{60}\!\left(-2\sqrt{15}\,z_1 - 3\sqrt{10}\,z_2
        - 5\sqrt{6}\,z_3 + 20\sqrt{3}\,z_4\right), \nonumber\\
  c_5 &= \tfrac{1}{60}\!\left(-2\sqrt{15}\,z_1 - 3\sqrt{10}\,z_2
        - 5\sqrt{6}\,z_3 - 10\sqrt{3}\,z_4 - 30\,z_5\right), \nonumber\\
  c_6 &= \tfrac{1}{60}\!\left(-2\sqrt{15}\,z_1 - 3\sqrt{10}\,z_2
        - 5\sqrt{6}\,z_3 - 10\sqrt{3}\,z_4 + 30\,z_5\right).
  \label{eq:6qubit-coeffs}
\end{align}
The logical state $\ket{\bar{1}}$ is determined analogously and depends on the same
parameters $z$; we omit it for brevity.

To confirm that every point $z \in \CC P^4$ yields a valid $((6,2,2))$ code, we
compute the weight enumerators~\cite{rainsWE}.
The $A$ and $B$ polynomials take the form
\begin{align}
  A(z) &= \bigl(1,\; 0,\; f^A_1(z),\; 0,\; f^A_2(z),\; 0,\; f^A_3(z)\bigr), \\
  B(z) &= \bigl(1,\; 0,\; f^B_1(z),\; 8,\; f^B_2(z),\; 24,\; f^B_3(z)\bigr),
\end{align}
where the entries list coefficients at weights $0,1,\ldots,6$ and
$f^A_i, f^B_i$ are explicit rational functions of $z$.
The vanishing of the weight-$1$ entries in both $A$ and $B$ confirms distance $2$
for every $z \in \CC P^4$.

One can further verify that $f^A_1(z) \neq f^B_1(z)$ for all $z$, establishing
that no member of this family achieves distance $3$: there is no hidden $((6,2,3))$
code within the moduli space.

\section{Further Realizations of $\lcode\irrep{5},2,2\rcode$}
\label{app:further-realizations}

The following subsections illustrate additional extrinsic realizations of the intrinsic
code $\lcode\irrep{5},2,2\rcode$ in continuous-variable and molecular settings.
In each case the embedding is $\SU(2)$-equivariant, so the Schur bootstrap
(Lemma~\ref{lem:SchurBootstrap}) guarantees detection of all errors transforming in
$\irrep{1}$ or $\irrep{3}$, including first-order rotational errors and first-order
momentum kicks.
Because $\irrep{5}$ is odd-dimensional, it appears in Hilbert spaces decomposing under
$\SO(3)$; the even-dimensional irrep $\irrep{14}$ of Example~2 does not appear in
purely rotational $\SO(3)$ decompositions, so none of the molecular realizations below
extend to that code without incorporating electronic or nuclear spin.

\subsection{Linear Rotor: AE Codes}
\label{app:AE}

A heterogeneous diatomic molecule (equivalently, a spinless particle on the sphere)
has rotational Hilbert space $\HH = L^2(S^2)$, which decomposes under $\SO(3)$ as
\begin{equation}
  L^2(S^2) = \bigoplus_{\ell \geq 0} \irrep{(2\ell+1)},
\end{equation}
with each irrep spanned by spherical harmonics $Y^\ell_m$.
The intrinsic code $\lcode\irrep{5},2,2\rcode$ resides in the $\ell=2$ sector via the
embedding $\emb:\ket{k}\mapsto Y^2_{2-k}$, giving codewords
\begin{align}
  \ket{\bar{0}} &= \tfrac{1}{\sqrt{2}}(Y^2_{-2}+Y^2_{2}), &
  \ket{\bar{1}} &= Y^2_{0}.
\end{align}
By the Schur bootstrap, this code detects all first-order rotational errors
$(J_\pm, J_z)$ and first-order momentum kicks $\widehat{\mathcal{Y}}^{L=1}_M$, both
of which transform in $\irrep{3}$.
More generally, an intrinsic code of depth $\mathsf{d}$ detects any combination of
rotational and momentum-kick errors whose combined adjoint order is less than
$\mathsf{d}$.
This is an example of an absorption-emission ($\mathcal{AE}$) code~\cite{Shubham},
a special case of the linear rotor codes of~\cite{ACP}.
A related bootstrap argument in this setting was examined in~\cite{AEbarg}.

\subsection{Rigid Rotor}
\label{app:rigid-rotor}

An asymmetric polyatomic molecule has rotational Hilbert space
$\HH = L^2(\SO(3))$, which by the Peter--Weyl theorem decomposes as
\begin{equation}
  L^2(\SO(3)) = \bigoplus_{\ell \geq 0} \irrep{(2\ell+1)} \otimes \irrep{(2\ell+1)}.
\end{equation}
The left quantum number $m$ is the space-fixed (lab-frame) projection and the right
quantum number $k$ is the body-fixed projection, with basis
$\ket{\begin{smallmatrix}\ell\\ mk\end{smallmatrix}}$.
There are two independent $\SU(2)$ actions: left rotations $\SU(2)_L$ (affecting $m$,
generated by $J_i^L$) and right rotations $\SU(2)_R$ (affecting $k$, generated by
$J_i^R$), along with orientation-operator errors $\widehat{D}^L_{MK}$ corresponding
to Wigner-$D$ matrices.

Unlike the linear rotor, the $\ell=2$ subspace $\irrep{5}_L\otimes\irrep{5}_R$
contains $\irrep{5}$ with multiplicity $5$, yielding two qualitatively distinct
families of embeddings.

\emph{Left and right sector embeddings.}
Projecting onto the left sector gives a $\CC P^4$ family of embeddings into
$\irrep{5}_L \subset \irrep{5}_L\otimes\irrep{5}_R$, and analogously for the right
sector.
Left-sector codes automatically protect against all right rotations (and vice versa),
and by the Schur bootstrap also satisfy the KL condition for first-order left
rotations $J_i^L$ and first-order momentum kicks $\widehat{D}^{L=1}_{MK}$, since
both transform in $\irrep{3}$ of $\SU(2)_L$.

\emph{Diagonal embedding.}
For any $\ell \geq 1$, the subspace $\irrep{(2\ell+1)}\otimes\irrep{(2\ell+1)}$
contains exactly one copy of $\irrep{5}$ under the diagonal $\SU(2)$ action.
For $\ell=1$ the unique diagonal embedding is
\begin{align*}
    \ket{0} &\mapsto \ket*{\smqty{1 \hfill \\  \smallminus 1, \smallminus 1}} \\
    \ket{1} &\mapsto \tfrac{1}{\sqrt{2}}\qty(\ket*{\smqty{1 \hfill \\ 0, \smallminus 1}} + \ket*{\smqty{1 \hfill \\ \smallminus 1, 0}} )\\
    \ket{2} &\mapsto \tfrac{1}{\sqrt{6}}\qty(\ket*{\smqty{1 \hfill \\ 1, \smallminus 1}} + 2\ket*{\smqty{1 \hfill \\ 0, 0}} + \ket*{\smqty{1 \hfill \\ \smallminus 1, 1}} )\\
    \ket{3} &\mapsto \tfrac{1}{\sqrt{2}}\qty(\ket*{\smqty{1 \hfill \\ 0,  1}} + \ket*{\smqty{1 \hfill \\  1, 0}} )\\
    \ket{4} &\mapsto \ket*{\smqty{1 \hfill \\ 1, 1}}. \numberthis
\end{align*}
giving codewords
\begin{align*}
    \ket{\overline{0}} &= \tfrac{1}{\sqrt{2}}\qty(\ket*{\smqty{1 \hfill \\  \smallminus 1, \smallminus 1}} + \ket*{\smqty{1 \hfill \\   1,  1}}) \\ \ket{\overline{1}} &= \tfrac{1}{\sqrt{6}}\qty(\ket*{\smqty{1 \hfill \\ 1, \smallminus 1}} + 2\ket*{\smqty{1 \hfill \\ 0, 0}} + \ket*{\smqty{1 \hfill \\ \smallminus 1, 1}} ). \numberthis
\end{align*}
This code protects against both first-order left and right rotational errors $J_i^L$
and $J_i^R$.
For the orientation errors $\widehat{D}^L_{MK}$, which transform in the reducible
representation $\irrep{(2L+1)}\otimes\irrep{(2L+1)}$, the code detects any
irreducible component transforming as $\irrep{3}$; there is precisely one such copy
for each $L \geq 1$.

\subsection{Molecular Codes}
\label{app:molecular}

A molecule with point-group symmetry $K \subset \SO(3)$ has rotational Hilbert space
$\HH = L^2(\SO(3)/K)$, which decomposes as
\begin{equation}
  L^2(\SO(3)/K) = \bigoplus_{\ell \geq 0} \irrep{(2\ell+1)} \otimes M_\ell^K,
\end{equation}
where $M_\ell^K$ is the $K$-fixed subspace of the body-fixed irrep
$\irrep{(2\ell+1)}$, with $\dim(M_\ell^K)$ giving the multiplicity of the
space-fixed irrep.
Modding out the right side preserves all left rotational errors while restricting
right rotational errors and momentum kicks by selection rules.

Sulfur dioxide has symmetry group $K = C_2$.
Taking each oxygen to have nuclear spin $0$, one finds $M_\ell^{C_2}$ consists of
even-$k$ values, so the multiplicity of $\irrep{5}$ in $L^2(\SO(3)/C_2)$ at $\ell=2$
is $3$.
The Schur bootstrap therefore yields a $\CC P^2$ family of extrinsic codes with
codewords
\begin{align*}
    \ket{\overline{0}} &= \sum_{\kappa \in \{-2,0,2\} }  \tfrac{z_\kappa}{\sqrt{2}}\qty( \ket*{\smqty{2\hfill\\-2,\kappa}} + \ket*{\smqty{2\hfill\\2,\kappa}}) \\
    \ket{\overline{1}} &= \sum_{\kappa \in \{-2,0,2\} } z_\kappa \ket*{\smqty{2\hfill\\0,\kappa}}, \numberthis
\end{align*}
where $z = [z_{-2}:z_0:z_2] \in \CC P^2$.
Each code detects all first-order left rotations (transforming as $\irrep{3}$),
protects against any right rotation, and detects all $\widehat{D}^{L=1}_{MK}$ errors.

The buckyball $\ce{{}^{12}C_{60}}$ provides a more striking example: its icosahedral
symmetry $K = I$ produces a Hilbert space $L^2(\SO(3)/I)$ in which codes within the
lowest nontrivial irrep $\irrep{13}$ can encode both a qubit and a qutrit with
first-order error protection.
The explicit constructions are given in the next appendix.

\subsection{Landau Level Codes}
\label{app:landau}

A charged particle on a sphere $S^2$ of radius $R$ in the field of a magnetic
monopole of charge $B R^2$ (with $qB > 0$) has quantized monopole strength
$j = qBR^2/\hbar$ (Dirac quantization) and Hilbert space~\cite{jacobsonsphere}
\begin{equation}
  \HH = \mathrm{Ind}_{\U(1)}^{\SU(2)}\,\chi_j
  = \bigoplus_{\ell \geq j} \irrep{(2\ell+1)},
\end{equation}
where $\ell$ and $j$ have the same integrality.
Each irrep $\irrep{(2\ell+1)}$ is a Landau level spanned by monopole spherical
harmonics ${}_jY^\ell_m$~\cite{monopoleharmonics}.
Setting $j = 0$ recovers the linear rotor $L^2(S^2)$.

For integral $j \leq 2$, the irrep $\irrep{5}$ appears once, giving the extrinsic code
\begin{align}
  \ket{\bar{0}} &= \tfrac{1}{\sqrt{2}}\bigl({}_jY^2_{-2}+{}_jY^2_2\bigr), &
  \ket{\bar{1}} &= {}_jY^2_0,
\end{align}
for $j \in \{0,1,2\}$.
Each realization protects against first-order rotations and first-order momentum kicks
${}_j\widehat{Y}^{L=1}_M$, both in $\irrep{3}$.

In the lowest Landau level limit (large $B$, small $R$, fixed $j$), the Landau levels
become energetically isolated and the effective Hilbert space collapses to
$\irrep{(2j+1)}$.
For $j=2$ this gives the spin-$2$ intrinsic code space itself, with only rotational
errors surviving; the code then protects against all first-order rotations.
A related construction was considered in~\cite{YaleFan}.

More generally, the Hilbert spaces in this appendix are all of the form
$\HH = \mathrm{Ind}_K^G\,\chi$ for a subgroup $K \subset G$ and a $K$-irrep $\chi$.
The Schur bootstrap applies uniformly across all such induced-representation Hilbert
spaces: any equivariant embedding of an intrinsic code into $\HH$ inherits the same
symmetry-resolved KL guarantees, with the multiplicity of the target irrep in $\HH$
determining the dimension of the resulting moduli space of extrinsic realizations.

\section{Buckyball Codes}
\label{app:buckyball}

Buckminsterfullerene $\ce{{}^{12}C_{60}}$ consists of 60 carbon atoms each with
nuclear spin $0$, so its rotational Hilbert space is $\HH = L^2(\SO(3)/I)$, where
$I \subset \SO(3)$ is the icosahedral group of order $60$.
This decomposes under $\SU(2)$ as
\begin{equation}
  L^2(\SO(3)/I) \;=\; \irrep{1} \oplus \irrep{13} \oplus \irrep{21} \oplus
  \irrep{25} \oplus \irrep{31} \oplus \cdots,
\end{equation}
where the smallest nontrivial irrep $\irrep{13}$ corresponds to the lowest nontrivial
rotational energy level.
Intrinsic codes within $\irrep{13}$ therefore yield, via the Schur bootstrap, the
first quantum error-correcting codes residing entirely within the lowest nontrivial
rotational level of the buckyball.

\subsection*{Qutrit Code}

There exists a $\lcode\irrep{13},3,2\rcode$ intrinsic $\SU(2)$ code with codewords
\begin{align*}
    \ket{\overline{0}} &= - \sqrt{\tfrac{21}{80}} |1\rangle +\tfrac{|3\rangle }{\sqrt{10}}+ \sqrt{\tfrac{63}{440}} |5\rangle - \sqrt{\tfrac{63}{176}}
   |9\rangle -\sqrt{\tfrac{3}{22}} |\text{11}\rangle \\
   \ket{\overline{1}} &= \tfrac{|0\rangle }{4 \sqrt{11}}+\sqrt{\tfrac{21}{55}} |2\rangle -\tfrac{3 |4\rangle }{4 \sqrt{5}}+\tfrac{3 |8\rangle }{4 \sqrt{5}}-\sqrt{\tfrac{21}{55}}
   |\text{10}\rangle -\tfrac{|\text{12}\rangle }{4 \sqrt{11}} \\
   \ket{\overline{2}} &= \sqrt{\tfrac{3}{22}} |1\rangle + \sqrt{\tfrac{63}{176}} |3\rangle - \sqrt{\tfrac{63}{440}} |7\rangle -\tfrac{|9\rangle
   }{\sqrt{10}}+ \sqrt{\tfrac{21}{80}} |\text{11}\rangle. \numberthis
\end{align*}
This code is covariant under the tetrahedral group $T$ (order $12$), consistent with
the intrinsic Eastin--Knill theorem (Lemma~\ref{lem:IntrinsicEK}), which requires the
logical symmetry group to be finite.

Via the Schur bootstrap, this yields an extrinsic qutrit code within the lowest
nontrivial energy level of the buckyball.
It detects all first-order left rotations (transforming as $\irrep{3}$) and protects
against any right rotation, since the code resides entirely in the left sector.
Notably, the momentum-kick operators $\widehat{D}^L_{MK}$ are automatically protected
up to and including fifth order due to the selection rules of the icosahedral Hilbert
space --- a property of $\HH$ rather than of the code itself.
Any gate in $T$ can be implemented by physical left rotations of the molecule.

\subsection*{Qubit Code}

There exists a $\lcode\irrep{13},2,3\rcode$ intrinsic $\SU(2)$ code, whose codeword
coefficients are determined by the unique positive real solution to the KL conditions
for this irrep and distance.
Defining the auxiliary quantities
\begin{align}
  v &= \frac{\sqrt[3]{611808967 + 9900\sqrt{3807314905}}}{1050}, \nonumber\\
  w &= v + \frac{104929}{1102500\,v} - \frac{11}{150}, \nonumber\\
  D &= \frac{49w^3}{66} + \frac{49w^2}{120} + w + 1,
\end{align}
the codewords are
\begin{align}
  \ket{\bar{0}} &= \tfrac{1}{\sqrt{D}}\!\left(
    \tfrac{7w^{3/2}}{\sqrt{66}}\ket{1} - \tfrac{7w}{2\sqrt{30}}\ket{4}
    + \sqrt{w}\,\ket{7} + \ket{10}\right), \nonumber\\
  \ket{\bar{1}} &= \tfrac{1}{\sqrt{D}}\!\left(
    \ket{2} + \sqrt{w}\,\ket{5} - \tfrac{7w}{2\sqrt{30}}\ket{8}
    + \tfrac{7w^{3/2}}{\sqrt{66}}\ket{11}\right).
\end{align}
This code is covariant under $S_3$ (the symmetric group on $3$ letters, order $6$),
again consistent with the finite logical symmetry required by the intrinsic
Eastin--Knill theorem.

Via the Schur bootstrap, this yields an extrinsic qubit code within the lowest
nontrivial energy level of the buckyball that \emph{corrects} any first-order error,
whether rotational or a momentum kick.
Any gate in $S_3$ is implementable by physical rotations of the molecule.

\subsection*{Remark}

The constructions above use only $\irrep{13}$, the lowest nontrivial level.
Codes of arbitrary dimension and distance can be obtained by working within higher
irreps $\irrep{21}$, $\irrep{25}$, $\irrep{31}$, \ldots, corresponding to higher
rotational energy levels of the buckyball.
The Schur bootstrap guarantees that any intrinsic code found within such an irrep
automatically yields a physically realizable buckyball code with the same
symmetry-resolved protection properties.

% \section{Related Work}

% Every permutation invariant (PI) quantum code can be thought of as an extrinsic realization of some intrinsic code (that is, the converse of the Schur-Bootstrap holds in this case). A proof is given in \cite{victornew}. 

% So naturally every PI code is just coming from an intrinsic code in disguise. But the benefit of the intrinsic picture is that you can then realize that same code in a variety of other settings. See \cite{RuskaiPRL,PI2,PI3,PIqudit1,us2} for many different examples of PI codes.

\section{Related Work}
\label{app:related}

A closely related recent work is that of Aydin, Albert, and Barg~\cite{victornew},
who develop a unified framework relating permutation-invariant (PI) codes,
generalized spin codes, and constant-excitation bosonic codes in the symmetric
$\SU(q)$ setting. Their paper makes explicit a striking correspondence among
these three code families and uses it in a constructive way, producing new
examples and transferring logical gates and error-correction properties between
the three models.

The present paper approaches the same phenomenon from a more representation-theoretic
point of view.  Rather than beginning with one of these concrete realizations,
we begin with the underlying representation itself.  In the symmetric case
$V=(n,0,\ldots,0)$, this abstract representation is precisely the generalized
spin space appearing in~\cite{victornew}, while the PI and bosonic models arise
as particular equivariant realizations of the same object.  From this viewpoint,
the relationship between PI, spin, and bosonic codes is not an isolated
three-way coincidence, but an instance of a more general principle: different
physical Hilbert spaces can carry the same representation, and therefore the
same intrinsic code.

This change in viewpoint has several consequences.  First, the Schur bootstrap
replaces model-specific transitions by a single uniform mechanism: once an
intrinsic code is fixed inside a representation $V$, every $G$-equivariant
realization of $V$ automatically inherits the same Knill--Laflamme relations.
In particular, the various arrows relating PI, spin, and bosonic codes become
manifestations of one representation-theoretic fact.  In this sense, the
intrinsic framework completes the correspondence picture of~\cite{victornew} by
showing that, whenever the same representation appears in PI, spin, and bosonic
Hilbert spaces, these are simply equivalent realizations of a single code.

Second, the scope is substantially broader.  The framework of~\cite{victornew}
is built around the symmetric $\SU(q)$ irreps $(n,0,\ldots,0)$, whereas the
intrinsic theory applies to arbitrary irreps of $\SU(q)$ and, more generally,
to representations of compact groups beyond $\SU(q)$.  Thus PI, spin, and
bosonic codes are only the first instances of the intrinsic picture.  The same
idea also produces realizations in settings such as linear and rigid rotors,
molecular rotational Hilbert spaces, and Landau levels, all of which fall
outside the three-model framework of~\cite{victornew}.

The Schur bootstrap should also be compared with our earlier Dicke bootstrap
construction~\cite{us2}, which established the passage from $\SU(2)$ spin codes
to permutation-invariant qubit codes.  That result appears here as a special
case.  The new proof is both cleaner and more revealing: instead of relying on
the special combinatorics of Dicke states, it follows directly from
$G$-equivariance and Schur's lemma, and it applies uniformly far beyond the
$\SU(2)$ permutation-invariant setting.

Taken together, these distinctions suggest the following perspective.
The work of~\cite{victornew} isolates and develops an important three-way
relationship inside the symmetric $\SU(q)$ world.  The intrinsic framework of
the present paper explains why that relationship exists, extends it to all
irreps and many more physical realizations, and shows that PI, spin, and
bosonic codes are three visible faces of a much broader representation-theoretic
structure.

\clearpage
\onecolumn
\section{Explicit Codewords for Intrinsic $  \lcode\irrep{27}, 5, 2\rcode_{\SU(3)}$
 code}
\label{app:explicitcodewords}

\begin{align*}
\ket{\overline{0}} &= \frac{|0\rangle }{2 \sqrt{2}}+\left(-\frac{1}{12 \sqrt{2}}+\frac{i}{4 \sqrt{6}}\right) |2\rangle +\left(-\frac{1}{12
   \sqrt{2}}+\frac{i}{4 \sqrt{6}}\right) |9\rangle +\left(\frac{5 \left(3 \sqrt{5}-1\right)}{48 \sqrt{3}}-\frac{5}{48} i
   \left(1+\sqrt{5}\right)\right) |12\rangle  \\
   &+\left(\frac{1}{16} \left(-5-\sqrt{5}\right)-\frac{1}{8} i \sqrt{\frac{5}{6} \left(7-3
   \sqrt{5}\right)}\right) |13\rangle +\left(\frac{\sqrt{5}-15}{24 \sqrt{3}}+\frac{1}{24} i \left(5+\sqrt{5}\right)\right)
   |14\rangle +\left(-\frac{1}{4 \sqrt{2}}+\frac{1}{4} i \sqrt{\frac{5}{6}}\right) |17\rangle  \\
   &+\left(-\frac{1}{4
   \sqrt{2}}-\frac{1}{4} i \sqrt{\frac{5}{6}}\right) |24\rangle +\left(\frac{1}{6 \sqrt{2}}-\frac{i}{2 \sqrt{6}}\right) |26\rangle \\
\ket{\overline{1}} &= \frac{|1\rangle }{\sqrt{3}}+\left(\frac{1}{48} \left(1-3 \sqrt{5}\right)-\frac{1}{8} i \sqrt{\frac{1}{6}
   \left(3+\sqrt{5}\right)}\right) |10\rangle +\left(\frac{1}{48} \left(\sqrt{5}-15\right)-\frac{1}{8} i \sqrt{\frac{5}{6}
   \left(3+\sqrt{5}\right)}\right) |11\rangle \\
   &+\left(\frac{1}{48} \left(1+3 \sqrt{5}\right)+\frac{1}{48} i
   \left(\sqrt{3}-\sqrt{15}\right)\right) |15\rangle +\left(\frac{1}{48} \left(15+\sqrt{5}\right)+\frac{i
   \left(\sqrt{5}-5\right)}{16 \sqrt{3}}\right) |16\rangle +\left(-\frac{7}{8 \sqrt{3}}-\frac{i \sqrt{5}}{8}\right) |25\rangle \\
\ket{\overline{2}} &= \frac{|3\rangle }{\sqrt{3}}+\left(\frac{1}{8} \sqrt{\frac{23}{3}+\sqrt{5}}+\frac{1}{8} i \sqrt{3-\sqrt{5}}\right) |6\rangle
   +\left(\frac{1}{48} \left(\sqrt{6}-3 \sqrt{30}\right)+\frac{1}{8} i \sqrt{3+\sqrt{5}}\right) |19\rangle +\left(-\frac{7}{8
   \sqrt{3}}+\frac{i \sqrt{5}}{8}\right) |23\rangle \\
\ket{\overline{3}} &= \frac{|4\rangle }{6}-\frac{\sqrt{5} |5\rangle }{6}+\left(\frac{3 \sqrt{5}-1}{8 \sqrt{3}}+\frac{1}{8} i
   \left(1+\sqrt{5}\right)\right) |8\rangle +\left(\frac{1}{24} \left(\sqrt{3}+3 \sqrt{15}\right)+\frac{1}{8} i
   \left(\sqrt{5}-1\right)\right) |18\rangle +\left(\frac{1}{12}+\frac{i}{4 \sqrt{3}}\right) |21\rangle \\
   & +\left(-\frac{\sqrt{5}}{12}-\frac{1}{4} i \sqrt{\frac{5}{3}}\right) |22\rangle \\
\ket{\overline{4}} &= \left(\frac{\sqrt{\frac{3}{2}}}{4}+\frac{1}{12} i \sqrt{\frac{5}{2}}\right) |2\rangle
   +\left(-\frac{\sqrt{\frac{3}{2}}}{4}+\frac{1}{12} i \sqrt{\frac{5}{2}}\right) |9\rangle +\left(\frac{5}{48}
   \left(1+\sqrt{5}\right)+\frac{5 i \left(\sqrt{5}-3\right)}{48 \sqrt{3}}\right) |12\rangle \\
   &+\left(-\frac{\sqrt{5}-15}{16
   \sqrt{3}}+\frac{1}{16} i \left(5+\sqrt{5}\right)\right) |13\rangle +\left(\frac{1}{24} \left(-5-\sqrt{5}\right)+\frac{1}{24} i
   \left(\sqrt{15}-\frac{5}{\sqrt{3}}\right)\right) |14\rangle +\left(-\frac{1}{4 \sqrt{6}}-\frac{i}{4 \sqrt{2}}\right) |17\rangle \\
   &
   +\left(\frac{1}{4 \sqrt{6}}+\frac{i}{4 \sqrt{2}}\right) |24\rangle -\frac{1}{6} i \sqrt{\frac{5}{2}} |26\rangle
\end{align*}

\end{document}